\documentclass[aps,pre,showpacs,amsmath]{revtex4}
\usepackage{amsbsy,amsmath,amssymb}
\usepackage{graphicx}% Include figure files
\begin{document}
%\draft

\title{Anomalous scaling of passively advected
magnetic field in the presence of strong anisotropy}

\author{
M. Hnatich$^{1}$, J. Honkonen$^{2}$, M. Jurcisin$^{1,3}$, A.
Mazzino$^{4}$, and S.~Sprinc$^{1}$}

\affiliation{
$^{1}$  Institute of Experimental Physics, Slovak Academy of Sciences,
Watsonova 47, 043 53 Ko\v{s}ice, Slovakia \\
$^{2}$  Department of Physical Sciences, Gustav H\"allstr\"omin katu 2, 00014
University of Helsinki and National Defence College, P.0. Box 7, 00861 Helsinki, Finland\\
$^{3}$  Bogolyubov
Laboratory of Theoretical Physics, Joint Institute for Nuclear
Research,
141 980 Dubna, Moscow Region, Russian Federation \\
$^{4}$ INFM-Department of Physics, University of Genova, I-16146 Genova, Italy}

\begin{abstract}
Inertial-range scaling behavior of high-order (up to order $N=51$) structure functions
of a passively advected vector field has been analyzed in the framework of the rapid-change
model with strong small-scale anisotropy with the aid of the renormalization
group and the operator-product expansion. It has been shown that
in inertial range the leading terms of the structure functions are coordinate independent,
but powerlike corrections appear with the same anomalous scaling exponents as for the passively advected
scalar field. These exponents depend on anisotropy parameters in such a way that a
specific hierarchy related to the degree of anisotropy is
observed. Deviations from power-law behavior like oscillations or
logarithmic behavior in the corrections to structure functions have not been found.
\end{abstract}

\pacs{PACS numbers: 47.10.+g, 47.27.$-$i, 05.10.Cc}

%\date{}
\maketitle

\section{Introduction}
\label{sec:Intro}

Justification of the basic principles of the Kolmogorov-Obukhov
(KO) phenomenological theory \cite{MoninYaglom,Orszag,Frisch} and
the investigation of possible deviations from its conclusions
within the framework of a microscopic model is one of the main
tasks in the theory of the fully developed turbulence and related
models, e.g. stochastic magnetohydrodynamics (MHD).

According to the KO theory, the following single-time structure functions
in the inertial range ($r_d \ll r \ll r_l$)
\begin{equation}
S_N(r)\equiv \langle [v_r({\bf x}, t)-v_r({\bf x^{\prime}}, t)]^N \rangle,
\,\,\,\, r=|{\bf x}-{\bf x^{\prime}}|
\label{stru}
\end{equation}
are independent of both the external (integral) scale $r_l$ and
internal (viscous) scale $r_d$, the latter being tantamount
to independence of viscosity.
These requirements are the famous first
and second hypotheses of Kolmogorov, respectively.
In Eq. (\ref{stru}) $v_r$ denotes the
component of the velocity field directed along the separation vector ${\bf
r}={\bf x}-{\bf x^{\prime}}$. Dimensional arguments then
determine the scale-invariant form of the structure functions (\ref{stru}) as
\[
S_N(r)=const \times (\bar{\epsilon} r)^{N/3}\,,
\]
where $\bar{\epsilon}$ is the mean dissipation rate \cite{MoninYaglom,Orszag,Frisch}.

On the other hand, both theoretical and experimental results reveal
some deviations from the KO theory
\cite{MoninYaglom,AnSaHu82},
viz. contradiction with the first Kolmogorov hypothesis.
For the structure functions (\ref{stru}) it means that they have
to be modified in the following way
\begin{equation}
S_N(r)=(\bar{\epsilon} r)^{N/3} \zeta_N(r/r_l)\,,
\label{basic1}
\end{equation}
where $\zeta_N(r/r_l)$ are scaling functions with powerlike
behavior in the asymptotic region $r/r_l \ll 1$
\begin{equation}
\zeta_N(r/r_l) \simeq const \times \left({r\over r_l}\right)^{q_N}.
\label{basic2}
\end{equation}
The singular dependence of the structure functions on $r_l$ in the
limit $r_l \rightarrow \infty$ together with nonlinearity of the
exponents $q_N$ as functions of $N$ is called "anomalous scaling".
Theoretical explanation of such behavior is based on strongly
developed fluctuations of the dissipation rate, i.e.
intermittency \cite{MoninYaglom,Frisch}.

A suitable and also powerful method of studying self-similar
scaling behavior is that of the renormalization group (RG)
\cite{Zinn,Vasilev,AdAnVa96,AdAnVa99}. In the theory of critical
phenomena it successfully explains the origin of the critical
scaling. This technique is also applicable to the theory of
turbulence, see Refs. \cite{Vasilev,AdAnVa99,deDoMa79} and references
therein.

The traditional approach to the description of fully developed
turbulence is based on the stochastic Navier-Stokes equation
\cite{wyld61}. Within the RG method the second Kolmogorov
hypothesis (independence of the viscous length $r_d$) was proved
for structure functions (\ref{stru}) for a variety of
realistic random forces, (see Ref. \cite{AdAnVa99}). This fact leads to
the existence of the IR scaling ($r \gg r_d$) with
definite "critical dimensions"
\begin{equation}
\Delta[S_N]=-{N\over 3}
\label{critdim2}
\end{equation}
with the customary convention $\Delta[r]=-1$ for scaling dimensions.
In the framework of the RG approach, critical dimensions like
(\ref{critdim2}) arise as coefficients in the differential RG
equations. On the other hand, it is not possible to
infer the form of the scaling functions
$\zeta_N(r/r_l)$ in Eq.\,(\ref{basic1}) from
the RG equations.

The standard way to investigate the dependence of the scaling
functions on the argument $r/r_l$ in the limit case $r/r_l
\rightarrow 0$ is the utilization of the
operator-product expansion (OPE), see Refs. \cite{Zinn,Vasilev,AdAnVa99}. The OPE
leads to the following representation of scaling
functions (\ref{basic2})
\begin{equation}
\zeta_N(r/r_l)=\sum_{F} C_{F}\,
\left({r\over r_l}\right)^{\Delta_{F}},\,\,\,r/r_l\rightarrow 0\,,
\label{strufun}
\end{equation}
where summation over all possible composite operators $F$
(i.e. products of fields and their derivatives) is
implied (for details, see Sec. \ref{sec:OPE}), $\Delta_{F}$ are their
critical dimensions, and  coefficients $C_{F}$ are regular
functions of $r_l^{-1}$.

Contrary to the theory of critical phenomena, where there are
physical reasons to believe that all relevant
composite operators have positive critical dimensions ($\Delta_{F}>0$)
for physical values of parameters [this is why the leading term in
the expansion (\ref{strufun}) is given by the trivial operator $F=1$
($\Delta_1=0$) in that case], in the theory of fully developed turbulence based
on the stochastic Navier-Stokes equation critical dimensions
of many composite operators are definitely negative for physical values of
parameters.
Existence of these "dangerous operators" leads to singular behavior of structure
functions in the limit $r/r_l \rightarrow 0$ \cite{AdAnVa99}. In the stochastic
Navier-Stokes model dangerous operators enter into the OPE in the
form of infinite families with the spectrum of critical dimensions
unbounded from below, and a nontrivial problem
of the summation of their contributions arises. This is an unsolved
problem of the theory.

In a situation where there are difficulties to study the anomalous
scaling in the stochastic Navier-Stokes model (this applies to
the stochastic MHD as well) it does not seem to be
unreasonable to consider simpler
models, which have features similar to real turbulent flow, and,
on the other hand, are easier for investigation. An important role in
this study was played by the model of passive advection of a
scalar quantity (temperature or concentration of the tracer) by a
uncorrelated-in-time Gaussian velocity field \cite{kr1}. Models
of passively advected vector fields \cite{adz11}
are straightforward generalizations
of the model of the passive advection of a scalar field.

In the present work the spatial structure of correlations of
fluctuations of the magnetic (vector) field ${\bf b}$ in a given turbulent
fluid in the framework of the kinematic MHD Kazantsev-Kraichnan model
(KMHD) is studied. These fluctuations are generated stochastically by
a Gaussian random emf and a white in time and anisotropic
self-similar in space Gaussian drift. The main goal is the calculation of
the anomalous exponents as functions of the anisotropy parameters of
the drift. From the mathematical point of view the present
model is found to be similar to the model
of a passive scalar quantity advected by a Gaussian strongly anisotropic
velocity field \cite{AdAnHnNo00} in that the fluctuation contributions
to the critical dimensions $\Delta_F$ of the OPE representation (\ref{strufun})
of the structure functions
coincide in both cases and thus the hierarchical dependence on the
degree of anisotropy is also the same. Here, numerical calculation of the critical dimensions
$\Delta_F$ in the one-loop approximation has been extended to dimensions related to
structure functions of order $N=51$ to explore possible
departures from powerlike asymptotic behavior.
However, contrary to the scalar case, in the inertial range the leading terms of
the structure functions of the magnetic field themselves are shown to be coordinate independent
with powerlike corrections whose exponents are generated by the
calculated critical dimensions.

\section{Kinematic MHD Kazantsev-Kraichnan model}
\label{sec:KMHD}

Consider passive advection of a solenoidal magnetic field
${\bf b} \equiv {\bf b}({\bf x},t)$ in the framework of the KMHD
model described by the stochastic equation
\begin{equation}
\partial_t {\bf b}  =  \nu_0 \triangle {\bf b} - ({\bf v \cdot \nabla}) {\bf b}
+ ({\bf b \cdot \nabla}) {\bf v} + {\bf f},
\label{K-K}
\end{equation}
where $\partial_t\equiv \partial/\partial t$, $\triangle\equiv{\bf \nabla}^2$
is the Laplace
operator, $\nu_0$ is the coefficient of the magnetic diffusivity,
and ${\bf v} \equiv {\bf v} ({\bf x} ,t)$ is a random solenoidal
(owing to the incompressibility) velocity field. Thus, both ${\bf
v}$ and ${\bf b}$ are divergence-free vector fields: ${\bf \nabla}
\cdot {\bf v}={\bf \nabla} \cdot {\bf b}=0$. A transverse Gaussian
emf flux density ${\bf f} \equiv {\bf f} ({\bf x} ,t)$ with
zero mean and the correlation function
\begin{equation}
D_{ij}^f \equiv \langle f_i({\bf x},t) f_j({\bf x^{\prime}},t^{\prime}) \rangle=
\delta(t-t^{\prime})C_{ij}({\bf r}/L), \,\,\,\,\ {\bf r}={\bf x}-{\bf x^{\prime}}
\label{cor-b}
\end{equation}
is the source of the fluctuations of the
magnetic field ${\bf b}$. The parameter $L$ represents an integral
scale related to the stirring, and $C_{ij}$ is a
function finite in the limit $L \rightarrow \infty$. In the present
treatment its precise form is irrelevant, and with no loss of
generality, we take $C_{ij}(0)=1$ in what follows. The random
velocity field ${\bf v}$ obeys Gaussian statistics with zero
mean and the correlation function
\begin{equation}
D_{ij}^v ({\bf x},t)  \equiv \langle v_i ({\bf x}, t) v_j (0,0)
\rangle = \frac {D_0 \delta (t)} {(2 \pi)^d} \int d^d {\bf k}\frac
{e^{i {\bf k \cdot x}}\, T_{ij} ({\bf k})} {(k^2 +
r_l^{-2}
)^{d/2+\epsilon/2}}, \label{cor-v}
\end{equation}
where
$r_l$ is another integral scale. In general, the scale $r_l$ may be
different from the integral scale $L$, below we, however,
take $r_l \simeq L$. $D_0>0$ is an amplitude factor related to the
coupling constant $g_0$ of the model by the relation $D_0/\nu_0
\equiv g_0 \equiv \Lambda^{\epsilon}$, where $\Lambda$ is the
characteristic UV momentum scale, and $0<\epsilon<2$ is a free
parameter. Its "physical" value $\epsilon =4/3$ corresponds to
the Kolmogorov scaling of the velocity correlation function in developed
turbulence.  $d$ is the dimensionality of the coordinate space. In
the isotropic case, the second-rank tensor $T_{ij}({\bf k})$ in
Eq.\,(\ref{cor-v}) has the simple form of the ordinary transverse
projector: $T_{ij}({\bf k})=P_{ij}({\bf k})\equiv\delta_{ij}-k_i
k_j/k^2$.

In what follows we will be interested in the asymptotic
behavior of the structure functions $S_N(r)$ within the
inertial range [defined by the inequalities $r_d \ll r \ll r_l$,
where $r_d\simeq \Lambda^{-1}$ is an internal (viscous) scale],
which represent the equal-time correlations of the $N$th power
of the projection of the field ${\bf b}$ onto the direction along
the separation vector of two different space points ${\bf x}$ and ${\bf x^{\prime}}$
\begin{equation}
S_{N}(r)\equiv
\langle[b_r({\bf x},t) - b_r({\bf x'},t)]^{N}\rangle, \qquad
r \equiv | {\bf x} - {\bf x'} |.
\label{struc}
\end{equation}
Dimensional analysis yields
\begin{equation}
S_{N}(r)=\nu_0^{-N/2} r^{N} R_{N}(r/r_d,r/r_l),
\label{srtuc100}
\end{equation}
where $R_{N}$ are functions of dimensionless parameters.
When the random source field ${\bf f}$ and the velocity field
${\bf v}$ are uncorrelated, the odd functions $S_{2n+1}$ vanish, however.
The standard perturbation expansion (series in $g_0$) is ill suited
for calculation of structure functions (\ref{srtuc100})
in the limit $r/r_d \rightarrow \infty$ and $r/r_l \rightarrow 0$,
due to the singular behavior of the coefficients
of the expansion. Therefore, to find the correct
IR behavior it is necessary to sum the whole series. Such
a summation can be carried out within the field-theoretic
RG and OPE. A compact description of this procedure is presented in
Refs.\,\cite{AdAnVa98,AdAn98,An99} (see also
Ref.\cite{AdAnHnNo00}). Below we remind basic ideas and results referring to the
isotropic case for simplicity of notation.

The RG analysis can be divided into two main parts. First, the
UV renormalization of structure functions
(\ref{struc}) is carried out. As a
consequence of this the asymptotic behavior of
functions like (\ref{struc}) for $r/r_d \gg 1$ and arbitrary
but fixed $r/r_l$ is given by IR stable fixed point(s) of the
corresponding RG equations and for functions (\ref{struc})
the following asymptotic form is obtained
\[
S_{N}(r)\sim\nu_0^{-N/2} r^{N}\left({r\over r_d}\right)^{-\gamma_{N}^*}
{\xi}_{N}(r/r_l),\,\,\,\,r/r_d \gg 1\,.
\]
where the scaling functions ${\xi}_{N}(r/r_l)$ remain unknown.
In the standard language of the theory of critical phenomena (see
for example Ref. \cite{Zinn,Vasilev}), the critical dimensions
$\Delta[S_{N}]$ of the functions $S_{N}$ are given by the
relations $\Delta[S_{N}]=-N+\gamma_{N}^*$, where $\gamma_{N}^*$
are "anomalous dimensions".
The dimensions
$\Delta[S_{N}]$ are calculated as series in $\epsilon$, and this
is why the exponent $\epsilon$ here plays the role analogous
to the parameter $4-d$ in the RG theory of critical
phenomena. Another parallel is related to the parameter $r_l$
which is an analog of the correlation length $r_c$
\cite{Zinn,Vasilev}.

Second,
the small $r/r_l$ behavior of the functions ${\xi}_{N}(r/r_l)$
has to be estimated. This
may be done using the OPE, which leads to the following
asymptotic form in the limit $r/r_l \rightarrow 0$
\[
{\xi}_{N}(r/r_l)=\sum_{F} C_{F}(r/r_l) \left({r\over r_l}\right)^{\Delta_{F}}\,
\]
where $C_{F}(r/r_l)$ are coefficients regular in $r/r_l$. The summation
is implied over all possible renormalized scale-invariant
composite operators $F$,
and $\Delta_{F}$ are their critical dimensions.

In the limit $L/r\to \infty $ correlation function (\ref{cor-b})
of the random source field is uniform in space, which -- as usual
in stochastic models describing turbulence \cite{AdAnHnNo00,AdAnVa98,AdAn98,An99,AdAnVa96,AdAnVa99}
-- brings about composite operators with negative critical dimensions
(dangerous composite operators) in the asymptotic analysis.
Contributions of these dangerous operators to the OPE imply singular behavior of the scaling
functions in the limit $r/r_l \rightarrow 0$. The leading term is given by
the operator with the most negative critical dimension $\Delta_F$. In our
model the leading contributions to even structure functions $S_{N}$ are given by scalar operators
$F_{N}=(b_i b_i)^{N/2}$ with their critical dimensions $\Delta_{N}$, which
eventually determine the asymptotic behavior of the structure functions $S_{N}$
of the form
\begin{equation}
S_{N}(r)\propto \nu_0^{-N/2} r^{N}
\left({r\over r_d}\right)^{-\gamma_{N}^*} \left({r\over r_l}\right)^{\Delta_{N}}=
 \nu_0^{-N/2} r_l^{-\Delta_{N}}
r_d^{\gamma_{N}^*}\,,
\label{struc120}
\end{equation}
where $\Delta_{N}=-N+\gamma_{N}^*$. Calculation shows
that  the anomalous dimensions $\gamma_{N}^*$ as well as the
critical dimensions $\Delta_{N}$ in the model considered
in the one-loop approximation
are related to critical dimensions of composite
operators of a simpler model of passively advected scalar
field \cite{AdAnVa98}, viz. $\gamma_{N}^*$
are given by
\begin{equation}
\gamma_{N}^*={N\epsilon\over 2}-{N (N/2-1)\epsilon\over(d+2)} + O(\epsilon^2)\,.
\label{critdim}
\end{equation}
From relation (\ref{struc120}) it follows that in the inertial range
the structure functions are flat: $S_{N}(r)\sim const\,$! Taking into account the nonlinear dependence
on $N$ of the anomalous dimensions (\ref{critdim}) we see that the deviation from the Kolmogorov
scaling is extremely large in this case. It should be emphasized that this conclusion persists to all
orders in perturbation theory and is not an artefact of the present one-loop calculation.
Below it will be shown that these relations are stable
against small-scale anisotropy.

In the anisotropic case we will assume that the statistics of the velocity field is anisotropic
at all scales and replace the ordinary transverse
projection operator in Eq.\,(\ref{cor-v})
with the operator
\begin{equation}
T_{ij} ({\bf k})  =
\left(1 + \alpha_{10} \frac {{\bf n \cdot k}} {k^2}\right) P_{ij} ({\bf k}) +
\alpha_{20} n_s n_l P_{is} ({\bf k}) P_{jl} ({\bf k})\,,
\label{T-ij}
\end{equation}
where $P_{ij} ({\bf k})$ is the usual transverse projection
operator, the unit
vector ${\bf n}$ determines the distinguished direction, and
$\alpha_{10}$, $\alpha_{20}$ are parameters characterizing the anisotropy. The
positive definiteness of the correlation function (\ref{cor-v})
imposes the following
restrictions on their values: $\alpha_{10}\,,\alpha_{20}>-1$. The operator
(\ref{T-ij}) is a
special case of the general transverse structure that possesses
uniaxial anisotropy:
\begin{equation}
T_{ij}({\bf k})= a(\psi) P_{ij}({\bf k}) +
b(\psi) n_s n_l P_{is} ({\bf k}) P_{jl}({\bf k})\,,
\label{generalT}
\end{equation}
where $\psi$ denotes the angle between the vectors ${\bf n}$ and ${\bf k}$
(${\bf n \cdot k}=k \cos \psi$). Using Gegenbauer polynomials \cite{GradshtejnRyzhik} the
scalar functions in representation (\ref{generalT}) may be expressed in the form
\[
a(\psi)=\sum_{i=0}^{\infty} a_{i} P_{2i}(\cos \psi)\,,\,\,\,\,
b(\psi)=\sum_{i=0}^{\infty} b_i P_{2i}(\cos \psi)\,.
\]
For the case of passively advected scalar,
it was shown in Ref. \cite{AdAnHnNo00} that all main features of the general
model with the anisotropy structure represented by Eq.\,(\ref{generalT}) are included in the
simplified model with the special form of the
transverse operator given by Eq.\,(\ref{T-ij}).
The same argument applies for the present case of passively advected
vector field as well.

The uniaxial anisotropy projector (\ref{T-ij}) has already been widely used
in analyzes of the anisotropically driven Navier-Stokes equation,
MHD turbulence equations and passive advection equations
\cite{RuBa87}.
However, these studies were limited to the investigation of the
existence and stability of the fixed points with the subsequent calculation
of the critical dimensions of the basic quantities leaving the calculation
of the anomalous exponents in those models an open problem.

The strong small-scale anisotropy (\ref{generalT}) does not change the leading inertial-range term
of the structure functions (\ref{struc120}), but the anisotropy shows in the corrections to it.
Indeed, combining the results of multiplicative renormalization
and OPE in the manner sketched above for the isotropic case, we arrive at the
conclusion that the inertial-range asymptotics of
the structure functions of the passively advected vector field
$S_{N}$ is a constant independent of $r$ with growing
powerlike corrections effected by the small-scale anisotropy:
\[
S_{N}(r)\sim D_0^{-N/2} r_d^{N-N\epsilon/2} \left({r_d\over r_l}\right)^{\Delta_{N}}
\left[c_{N}+
\sum\limits_{p=1}^{N}c_{N,p}\left({r\over
r_l}\right)^{\Delta_{[N,p]}-\Delta_{N}}
+\sum\limits_{{M+K=N\atop p\le M\,, q\le K}}c_{M,p,q}
\left({r\over r_l}\right)^{\Delta_{[M,p]}+\Delta_{[K,q]}-\Delta_{N}}\right]\,,
\]
where the critical dimensions $\Delta_{N}$ are given by Eq.\,(\ref{critdim})
at the order $O(\epsilon)$, whereas the critical dimensions $\Delta_{[M,p]}$
will be defined and calculated below [Eq. (\ref{deltaff})].
This is similar to the asymptotic behavior of the structure functions
of the passive scalar advected by a {\em compressible} vector field \cite{AdAn98}.

\section{Field-Theoretic Formulation, Renormalization, and RG Analysis}
\label{sec:FTRG}

The stochastic problem (\ref{K-K})--(\ref{cor-v}) is equivalent to
the field-theoretic model of the set of the three fields
$\Phi=\{{\bf b^{\prime}}, {\bf b}, {\bf v}\}$ with the action
functional
\begin{equation}
S(\Phi)  \equiv {1\over 2}\,{\bf b}'D_b {\bf b}' + {\bf b}'[ -\partial_{t}
-({\bf v} \cdot \nabla) + \nu_0 \Delta ]{\bf b} + {\bf b}'({\bf b}
\cdot \nabla){\bf v}
 - {1\over 2}{\bf v} D_v^{-1} {\bf v},
 \label{action1}
\end{equation}
where ${\bf b}'$ is an auxiliary field (all required
integrations over space-time coordinates and summations over the
vector indices are implied). The first five terms in
Eq.\,(\ref{action1}) represent the De Dominicis-Janssen action
corresponding to the stochastic problem at fixed ${\bf v}$ (see,
e.g., Refs.\,\cite{MaSiRo73}), whereas the
last term represents the Gaussian averaging over ${\bf v}$. $D_b$
and $D_v$ are the correlation functions (\ref{cor-b}) and
(\ref{cor-v}), respectively.

In this field-theoretic language, the structure functions
(\ref{struc}) are defined as
\begin{equation}
S_{N}(r)\equiv \int {\mathcal D} \Phi  [b_r({\bf x},t) - b_r({\bf
x'},t)]^{N} e^{S(\Phi)}
\label{strukt}
\end{equation}
with the action $S(\Phi)$ defined above.

Action (\ref{action1}) is given in a form convenient for
application of the quantum-field perturbation analysis with the
standard Feynman-diagram technique. The quadratic part of the
action determines the matrix of bare propagators.
For the fields
${\bf b^{\prime}}$ and ${\bf b}$ the propagators in the
wave-vector-frequency representation are
\begin{align}
\langle b_i b_j^{\prime} \rangle_0 &= \langle b_j^{\prime} b_i
\rangle^*_0= \frac{P_{ij}({\bf k})}{-i\omega + \nu_0 k^2}\,,\nonumber \\
\langle b_i b_j \rangle_0 &= \frac{C_{ij}({\bf
k})}{\omega^2+\nu_0^2
k^4}\,, \label{propagators}\\
\langle b_i^{\prime} b_j^{\prime} \rangle_0 &= 0\,,\nonumber
\end{align}
where $C_{ij}({\bf k})$ is the Fourier transform of the function
$C_{ij}({\bf r}/L)$ from Eq. (\ref{cor-b}). The bare propagator of the
velocity field
$\langle {\bf v} {\bf v}\rangle_0 \equiv \langle
{\bf v} {\bf v}\rangle$ is defined by Eq.\,(\ref{cor-v}) with the
transverse projector given by Eq.\,(\ref{T-ij}).
The interaction in the model is  given  by the nonlinear terms
$- b^{\prime}_i ({\bf v}\cdot \nabla) b_i + b^{\prime}_i ({\bf
b}\cdot \nabla) v_i \equiv b^{\prime}_i V_{ijl} v_j b_l$ with
the vertex factor which in the wave-number-frequency representation
has the following form
\[
V_{ijl}=i(\delta_{ij} k_l -\delta_{il} k_j).
\]
With the use of the standard power counting
\cite{Zinn,Vasilev} (see also Ref.\,\cite{AdAnHnNo00}
for peculiarities of rapid-change passive advection models)
correlation functions with superficial UV divergences
may be identified. These are correlation functions containing
frequency-wave-vector integrals divergent in the limit
$\epsilon\to 0$ with divergences brought about by integration over
large wave numbers and correspondingly having non-negative
wave-number dimension. In the present
model superficial divergences exist only in the
one-particle-irreducible (1PI) Green function $\Gamma_{{\bf b^{\prime}}{\bf b}}$.
In the isotropic case this Green function
gives rise only to the renormalization of
the term $\nu_0{\bf b}' \Delta {\bf b}$
of action (\ref{action1}) and the
corresponding
UV divergences may be fully
absorbed in the proper redefinition of the existing parameters
$g_0$, $\nu_0$ so that all correlation functions calculated in terms
of the renormalized parameters $g$ and $\nu$ are UV finite.

When anisotropy is introduced, however, the situation becomes more complicated, because
the 1PI Green function $\Gamma_{{\bf b^{\prime}}{\bf b}}$
produces divergences corresponding to
the structure ${\bf b^{\prime}} ({\bf n}\nabla)^2 {\bf b}$ in the action of the model
[due to peculiarities of the rapid-change models \cite{AdAnVa98}
the term $({\bf b^{\prime}}{\bf n}) \triangle ({\bf b}{\bf n})$
possible on dimensional and symmetry grounds does not appear].
The term ${\bf b^{\prime}} ({\bf n}\nabla)^2 {\bf b}$
is not present in the
original unrenormalized action (\ref{action1}), but has to be added to the renormalized action,
therefore the model is not multiplicatively renormalizable. In such a case it is customary
to extend the original action (\ref{action1}) by including all terms needed
for the renormalization of the correlation functions and thus adding
new parameters. As a result the extended model is described
by a new action of the form:
\begin{equation}
S(\Phi)  \equiv {1\over 2}{\bf b'} D_b {\bf b'} + {\bf b}'[ -\partial_{t}
-({\bf v} \cdot \nabla) + \nu_0 \Delta + \chi_0\nu_0 ( {\bf n}
\cdot \nabla )^{2} ]{\bf b} + {\bf b'}({\bf b} \cdot \nabla){\bf
v}
 - {1\over 2}{\bf v} D_v^{-1} {\bf v},
 \label{action2}
\end{equation}
where a new unrenormalized parameter $\chi_0$ has been introduced.

Of course, the
bare propagators (\ref{propagators}) of the isotropic model are
modified and for the extended action (\ref{action2}) assume the form
\begin{align}
\langle b_i b_j^{\prime} \rangle_0 &= \langle b_j^{\prime} b_i
\rangle^*_0= \frac{P_{ij}({\bf k})}{-i\omega + \nu_0 k^2+\chi_0 \nu_0
({\bf n}\cdot {\bf k})^2}\,,\label{bbprime} \\
\langle b_i b_j \rangle_0 &= \frac{C_{ij}({\bf k})}{|-i
\omega+\nu_0
k^2 +\chi_0 \nu_0 ({\bf n}\cdot {\bf k})^2|^2}\,, \label{bb}\\
\langle b_i^{\prime} b_j^{\prime} \rangle_0 &= 0\,.\nonumber
\end{align}
After this modification all terms needed to
remove the divergences are present in action (\ref{action2}),
therefore the model becomes multiplicatively renormalizable allowing for the
standard RG analysis.
The corresponding renormalized action may be written down immediately:
\begin{equation}
S_R(\Phi)  \equiv {1\over 2}{\bf b'} D_b {\bf b'} + {\bf b}'[
-\partial_{t} -({\bf v} \cdot \nabla) + \nu Z_1 \Delta + \chi \nu
Z_2 ( {\bf n} \cdot \nabla )^{2} ]{\bf b} + {\bf b'}({\bf b} \cdot
\nabla){\bf v}
 - {1\over 2}{\bf v} D_v^{-1} {\bf v}\,.
 \label{action3}
\end{equation}
Here, $Z_1$ and $Z_2$ are the renormalization constants
in which the UV divergent parts of the 1PI response
function $\Gamma_{{\bf b^{\prime}}{\bf b}}$ are absorbed. The
renormalized action (\ref{action3}) leads to the multiplicative
renormalization of the parameters $\nu_0, g_0$ and $\chi_0$:
\[
\nu_0=\nu Z_{\nu},\,\,\,\, g_0=g \mu^{\epsilon} Z_g,\,\,\,\,
\chi_{0}=\chi Z_{\chi},
\]
where $\nu, g$, and $\chi$
are renormalized counterparts of the
bare parameters, and $\mu$ is a scale-setting parameter
with the same canonical dimension as the wave number.
The anisotropy parameters
$\alpha_{01}$ and $\alpha_{02}$ are not renormalized, therefore their renormalized
counterparts
$\alpha_{1}$ and $\alpha_{2}$ may be put equal to the unrenormalized parameters.
In what
follows, we will work in the minimal subtraction (MS) scheme, in which
-- in the
one-loop aproximation --
renormalization constants have the form $1+A/\epsilon$,
where the amplitude $A$ is a function of
$g,\chi,\alpha_1,\alpha_2$, and $d$, but independent of $\epsilon$.

\begin{figure}
\unitlength=1.00mm
\special{em:linewidth 0.4pt}
\linethickness{0.4pt}
\begin{picture}(96.67,8.74)
\put(43.67,2.72){\makebox(0,0)[ rc]  {$\Sigma_{b^{\prime}b}=$}}
\put(60.17,2.73){\oval(15.00,10.32)[ t]  }
\put(60.17,2.73){\oval(15.00,10.32)[ b]  }
\put(67.67,2.73){\line(1,0){7.00}}
\put(45.67,2.73){\line(1,0){7.00}}
\put(49.33,0.57){\line(0,1){4.30}}
\put(64.67,5.73){\line(0,1){3.01}}.
\end{picture}
\caption{\label{obr1}
The (exact) graphical expression for the self-energy operator $\Sigma_{b^{\prime}b}$ of the response function of the passive vector field.
The plain line denotes the bare
propagator (\ref{bb}), and the line with slash
(denoting the end corresponding to the arguments of the field ${\bf b^{\prime}}$) corresponds to
the bare propagator (\ref{bbprime}).
}
\end{figure}
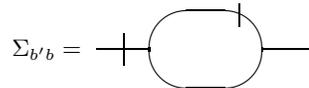

Identification of the unrenormalized action (\ref{action2}) with
the renormalized one (\ref{action3}) leads to the following
relations between the renormalization constants:
\[
Z_1=Z_{\nu},\,\,\,\,Z_2=Z_{\chi}Z_{\nu},\,\,\,\,Z_g=Z_{\nu}^{-1}\,.
\]
It has to be mentioned that the rapid-change models like (\ref{action2})
have the nice feature that in all multiloop diagrams of the
self-energy operator $\Sigma_{b^{\prime} b} $ closed circuits of the retarded
bare propagators $\langle {\bf b} {\bf b^{\prime}}\rangle_0$ are produced,
because the propagator $\langle{\bf v} {\bf v}\rangle_0$ is proportional
to the $\delta$ function in time.
As a result, the one-loop self energy operator $\Sigma_{b^{\prime} b} $
with the graphical notation of Fig. \ref{obr1} is exact.

The divergent part of the graph in Fig. \ref{obr1} is
\begin{equation}
\Sigma_{b^{\prime} b}(p)= - \frac{g \nu C_d}{2
d (d+2)\epsilon}
\left\{\left[(d-1)(d+2)+ \alpha_1(d+1)+\alpha_2\right]
p^2-(2\alpha_1-(d^2-2)\alpha_2)({\bf n}\cdot {\bf p})^2\right\}\,,
\label{sigmaa}
\end{equation}
where  $C_d \equiv S_d/(2\pi)^2$ and
$S_d=2\pi^{d/2}/\Gamma(d/2)$ is the area of the $d$-dimensional
sphere of unit radius.
The expression (\ref{sigmaa}) leads to a straightforward determination of the
renormalization constants $Z_{1}$ and $Z_{2}$:
\begin{align}
Z_1 &= 1-\frac{g C_d}{2 d (d+2) \epsilon} \left[(d-1)(d+2)+
\alpha_1(d+1)+\alpha_2 \right]
\,,\nonumber \\
Z_2 &= 1-\frac{g C_d}{2 d (d+2) \chi \epsilon}
\left[-2\alpha_1+(d^2-2)\alpha_2\right] \,.
\label{Z12}
\end{align}
With the proper choice of the renormalization constants this
renormalization procedure gives rise to UV-finite correlation functions
with separate space and time arguments. Independence of the original
unrenormalized model of the scale-setting parameter $\mu$ of the
renormalized model yields the RG differential equations for
the renormalized correlation functions of the fields, e.g.
\begin{equation}
({\mathcal D}_{\mu} + \beta_g \partial_g + \beta_{\chi}
\partial_{\chi} - \gamma_{\nu} {\mathcal D}_{\nu})
\langle {\bf b}({\bf x},t){\bf b}({\bf x}',t)\rangle_R =0\,
\label{rgeq1}
\end{equation}
with the definition ${\mathcal D}_x \equiv x \partial_x$ for any
variable $x$, and with the following definition of the RG functions (the $\beta$ functions and the
anomalous dimensions $\gamma$)
\[
\gamma_i \equiv \widetilde {\mathcal D}_{\mu} \ln Z_i
\]
for any renormalization constant $Z_i$, and
\[
\beta_g\equiv \widetilde {\mathcal D}_{\mu}
g=g(-\epsilon+\gamma_1)\,,\,\,\,\,\, \beta_{\chi}\equiv \widetilde
{\mathcal D}_{\mu} \chi=\chi(\gamma_1-\gamma_2)\,,
\]
where $\widetilde {\mathcal D}\equiv \mu \partial_{\mu}$ denotes
derivative with fixed bare parameters of the extended action (\ref{action2}).

For the anomalous dimensions $\gamma_{1}$ and $\gamma_{2}$ we obtain
from Eq. (\ref{Z12})
\begin{align}
\gamma_1 &=\frac{g C_d}{2 d (d+2)} \left[(d-1)(d+2)+
\alpha_1(d+1)+\alpha_2 \right] \,, \label{anomal1}\\
\gamma_2 &= 1-\frac{g C_d}{2 d (d+2) \chi}
\left[-2\alpha_1+(d^2-2)\alpha_2\right] \,. \label{anomal2}
\end{align}
It should be emphasized that both the renormalization
constants (\ref{Z12}) and the corresponding anomalous dimensions
(\ref{anomal1}) and (\ref{anomal2}) in the present model are exact, i.e., they have
no corrections of order $g^2$ or higher.

The fixed points ($g_*$, $\chi_*$) of the RG equations are defined by the system of two
equations
\[
\beta_g(g_*,\chi_*)=0\,,\qquad \beta_\chi(g_*,\chi_*)=0\,.
\]
The IR stability of a fixed point is
determined by the condition that real parts of the eigenvalues of the
matrix
\[
\omega=\left(
\renewcommand{\arraystretch}{1.5}
\begin{array}{cc}
\partial_g\beta_g&\partial_\chi\beta_g\\
\partial_g\beta_\chi&\partial_\chi\beta_\chi
\end{array}
\right)_{g=g_*\atop\chi=\chi_*}
\]
are positive.
Calculation shows that the RG equations have
only one non-trivial IR stable fixed point defined by expressions
\begin{align}
g_*&=\frac{2 d (d+2) \epsilon}{C_d\left[(d-1)(d+2)+
\alpha_1(d+1)+\alpha_2\right]}\,,\label{fixg} \\
\chi_*&=\frac{-2\alpha_1+(d^2-2)\alpha_2}{(d-1)(d+2)+
\alpha_1(d+1)+\alpha_2}. \label{fixchi}
\end{align}
Both eigenvalues of the stability matrix
$\omega$
are equal to $\epsilon$ at this fixed point, therefore, the IR
fixed point (\ref{fixg}), (\ref{fixchi})
is stable for $\epsilon>0$ and all values of the anisotropy
parameters $\alpha_1$ and $\alpha_2$.

Rather unexpectedly, the $\beta$ functions and, consequently, the fixed points
of the present model of
passively advected vector field are exactly the same as
in the model of passively advected scalar field
\cite{AdAnHnNo00}. In Sec. \ref{sec:OPE} it will be shown that
this similarity is extended to the anomalous scaling dimensions of the composite operators
in the OPE representation of the structure functions as well.

The fixed point (\ref{fixg}), (\ref{fixchi}) governs the behavior of solutions of
Eqs. (\ref{rgeq1})
and the like, and at large scales far from viscous length
$r\gg r_d$ at any fixed ratio $r/r_l$ yields the scaling
form
\begin{equation}
\label{bbscaling}
\langle {\bf b}({\bf x},t){\bf b}({\bf x}-{\bf r},t)\rangle
=D_0^{-1} r^{2-\epsilon}  \xi_2(r/r_l)\,,
\end{equation}
for the {\em unrenormalized} correlation function (we remind that due to the absence of
field renormalization renormalized and unrenormalized correlation functions are equal but expressed
in terms of different variables). It should be noted, however, that
the scaling function $\xi_2(r/r_l)$ in Eq. (\ref{bbscaling}) is not determined by the RG Eqs.
(\ref{rgeq1}).

This, however, is not enough to
find the asymptotic scaling behavior of the structure functions
(\ref{strukt}), because they are
linear combinations of correlation functions with coinciding arguments.
The latter
contain UV divergences additional to those included in the
renormalization constants (\ref{Z12}).
These additional divergences due to
composite operators
(products of fields and their derivatives with coinciding space and time
arguments)
can be dealt with in a manner similar to that applied to
the divergences in
the usual correlation functions \cite{Vasilev}.

\section{Renormalization and Critical Dimensions of Composite Operators}
\label{sec:OPE}

A composite operator is any product of
fields and their derivatives at
a single space-time point $x\equiv ({\bf x}, t)$, e.g.
$[{\bf b}^2(x)]^N$ and  $\left[\partial_i b_j(x) \partial_i b_j(x)\right]^N$.
Structure functions (\ref{strukt}) contain products of composite
operators at two separate space points. These composite operators are
integer powers of the field $b_r$. Usually, the aim of renormalization of composite
operators is to make UV finite all 1PI correlation functions with the insertion of
a composite operator $F$, i.e. quantities of the form
\[
\langle F\,\phi_{i_1}({\bf x}_1,t_1)
\cdots\phi_{i_m}({\bf x}_m,t_m)\rangle_{\rm 1PI}\,.
\]
Correlation functions with such insertions contain additional UV
divergences, which also may be removed by a suitable renormalization
procedure \cite{Zinn,Vasilev}. Composite operators mix in
renormalization, therefore an UV finite renormalized operator $F^R$
has the form $F^R=F+\sum\Delta F$, where the counterterms $\sum\Delta F$ are
a linear combination of the operator $F$ itself and, in general, other
unrenormalized operators required to make all correlation functions
-- generated by the renormalized action --
with the insertion of $F^R$ UV finite.
In this case homogeneous RG equations of the form (\ref{rgeq1})
may be obtained for certain linear combinations of
renormalized correlation functions with  composite-operator insertions.
Such linear combinations (basis operators $\mathcal{F}$, see below)
exhibit IR scaling with definite
critical dimensions $\Delta_{\mathcal{F}}$, whereas an arbitrary
renormalized composite operator may be expressed as a linear combination of these
basis operators.

The general procedure is the following
\cite{Zinn,Vasilev,AdAnVa96,AdAnVa99,AdAnVa98,AdAnHnNo00}: If
$\{F_{\alpha}\}$ is a closed set of composite operators (i.e.,
they are mixed only with each other in renormalization), then
the sets of renormalized and unrenormalized operators are
related through the matrix transformation:
\begin{align}
F_{\alpha}&=\sum_{\beta} Z_{\alpha \beta} F^R_{\beta}\,,\label{ZF}\\
\gamma_F&=Z_F^{-1} \widetilde{{\mathcal D}}_{\mu} Z_F\,,\label{anomal}
\end{align}
where $Z_F \equiv \{Z_{\alpha \beta}\}$ is the
renormalization matrix and $\gamma_F\equiv
\{\gamma_{\alpha \beta}\}$ is the
corresponding matrix of anomalous dimensions for this set of operators.
The renormalized composite operators obey the RG differential
equations
\[
({\mathcal D}_{\mu} + \beta_g \partial_g + \beta_{\chi}
\partial_{\chi} - \gamma_{\nu} {\mathcal D}_{\nu})
F^R_\alpha=-\sum\limits_\beta\gamma_{\alpha \beta}F^R_\beta\,,
\]
which give rise to the
matrix of critical dimensions $\Delta_F\equiv
\{\Delta_{\alpha \beta}\}$ of the form
\begin{equation}
\Delta_F=d^k_F-\Delta_t d^{\omega}_F +\gamma^*_F\,,\,\,\,\,\,\,
\Delta_t=-2+\epsilon\,,
\label{deltaf}
\end{equation}
where $d^k_F$ and $d^{\omega}_F$, respectively, are the diagonal matrices of
canonical wave-number and frequency
dimensions of the operators
(where the diagonal elements are sums of the corresponding
dimensions of the operators included in the composite operator)
and $\gamma^*_F$ is the matrix
of anomalous dimensions (\ref{anomal}) at the fixed point
(\ref{fixg}), (\ref{fixchi}).

Critical dimensions of the set $F\equiv \{F_{\alpha}\}$ are given
by the eigenvalues of the matrix $\Delta_F$. The basis operators
possessing definite critical dimensions are related to the
renormalized composite operators by the matrix transformation
\begin{equation}
\label{basisF}
\mathcal{F}_{\alpha}=\sum_{\beta}U_{\alpha \beta} F^R_{\beta}\,,
\end{equation}
where the matrix $U_F \equiv \{U_{\alpha \beta}\}$ is such that
the transformed matrix of critical dimensions
$\Delta_\mathcal{F}=U_F \Delta_F U^{-1}_F$ is diagonal.

The structure functions contain, however, quantities which
correspond to insertions of two composite operators.
Therefore, it would seem that
we would have to consider renormalization of products of two
composite operators as well, the aim being then to render
UV finite all 1PI correlation functions with two insertions of
composite operators. Superficially divergent correlation functions
with operator insertions are identified by power counting similar to
that of the basic renormalization. In the present
model such a power counting shows that insertion of products of composite operators of the structure
${\bf b}^m({\bf x},t){\bf b}^n({\bf x}',t)$ does not bring about any new superficial divergences
and it is thus sufficient to renormalize the composite operators
themselves only in order to make the structure functions UV finite.
Therefore, from the RG analysis of composite operators it follows
-- by virtue of relations (\ref{ZF}) and (\ref{basisF}) -- that
the structure function $S_N$ may be expressed as a functional average
of a quadratic form of basis operators:
\begin{equation}
\label{SNF}
S_N(r)=
\sum\limits_{\alpha,\beta}B_{\alpha\beta}
\left\langle\mathcal{F}_\alpha\left({\bf x}+{{ 1\over 2}}{\bf r},t\right)\mathcal{F}_\beta
\left({\bf x}-{1\over 2}{\bf r},t\right)\right\rangle_{\!\!R}
\end{equation}
with coefficients $B_{\alpha\beta}$ independent of spatial coordinates.
Each term in expression (\ref{SNF}) obeys the following
asymptotic form in the limit $r_d \ll r$, $r \lesssim r_l$
\begin{equation}
\label{basicF}
\left\langle\mathcal{F}_\alpha\left({\bf x}+{{ 1\over 2}}{\bf r},t\right)\mathcal{F}_\beta
\left({\bf x}-{1\over 2}{\bf r},t\right)\right\rangle_{\!\!R}\sim
D_0^{d^\omega_{\alpha}+d^\omega_{\beta}}
r^{-\Delta_{\alpha}-\Delta_{\beta}}
r_d^{\gamma^*_{\alpha}+\gamma^*_{\beta}}
\Xi_{\alpha\beta}\left({r\over r_l}\right)
\end{equation}
with the scaling functions $\Xi_{\alpha\beta}$ still to be
determined.

The physically interesting range of scales, however, is the inertial range,
specified by the inequalities $r_d \ll r\ll r_l$. The limit $r\ll r_l$
may be explored with the use of the OPE \cite{Zinn,Vasilev} as was
already discussed in Sec. \ref{sec:Intro}.
The basic statement of the OPE theory is that the equal-time product
of two renormalized composite operators can be represented in the form
\begin{equation}
F^R_\alpha\left({\bf x}+{{ 1\over 2}}{\bf r},t\right){{F}^R_\beta}\left({\bf x}-{{ 1\over 2}}{\bf r},t\right)
=\sum_{\gamma} C_{\alpha\beta\gamma} ({\bf r}) {F}^R_\gamma({\bf x},t),
\label{OPE}
\end{equation}
where the functions $C_{\alpha\beta\gamma}$  are the Wilson coefficients regular
in $1/r_l$, and ${F}^R_\gamma$ are renormalized local
composite operators which appear in the
formal Taylor expansion with respect to ${\bf r}$ together with all operators
that mix with them in renormalization. If these operators have
additional vector indices, they are contracted with the
corresponding indices of the coefficients $C_{\alpha\beta\gamma}$.

Without loss of generality we may take the expansion
on the right-hand side of Eq. (\ref{OPE}) in terms of the basis operators with
definite critical dimensions $\Delta_{\mathcal{F}}$. The renormalized
correlation function $\langle F^R_\alpha{{F}^R_\beta} \rangle_R$
is obtained by averaging Eq. (\ref{OPE}) with the weight
generated by the renormalized action, the quantities
$\langle \mathcal{F} \rangle_R$
appear now only on the right-hand side. Their asymptotic behavior
for $r/r_l\to 0$ is found from the corresponding RG equations
and is of the form $\langle \mathcal{F} \rangle \propto  r_l^{-\Delta_\mathcal{F}}$.
Comparison of the expression for a given function $\langle F^R_\alpha{F}^R_\beta \rangle_R$ in terms of
the IR scaling representation of correlation functions of the basis operators (\ref{basicF})
on one hand and the OPE representation brought about by relation (\ref{OPE}) on the other in the
limit $r_l\to \infty$ allows to find the asymptotic form of the scaling functions
$\Xi_{\alpha\beta}(r/r_l)$ in relation (\ref{basicF}).

The composite operators appearing in the expression
for the structure function $S_N$ are products of
integer powers of the field $b_r$ of the form
$b_r^{N-m}({\bf x},t)b_r^{m}({\bf x}',t)$.
Thus, at the leading order in ${\bf r}$ their OPE contains operators
of the closed set generated by the operator $b_r^{N}({\bf x},t)$.
Power counting and analysis of the structure of graphs shows
that this set of composite operators contains only operators consisting
of exactly $N$ components of the vector field ${\bf b}$ (i.e. no derivatives of the field
components appear).
Extracting the common scaling factor prescribed by the canonical dimensions
(\ref{deltaf})
of these operators the basis-operator decomposition
of any term of the structure function $S_N$ may be written as
\begin{equation}
\label{Sbb}
\left\langle b_r^{N-m}\left({\bf x}+{{ 1\over 2}}{\bf r},t\right)\,
b_r^{m}\left({\bf x}-{{ 1\over 2}}{\bf r},t\right)\right\rangle\sim
D_0^{-N/2}
r^{N(1-\epsilon/2)}\sum\limits_{\alpha,\beta,\gamma}
A_{\alpha\beta\gamma}({r/r_l})
\left({r\over r_d}\right)^{-\gamma^*_{_\alpha}-\gamma^*_{\beta}}
\left({r\over r_l}\right)^{\Delta_{\gamma}}
\end{equation}
where the coefficients $A_{\alpha\beta\gamma}(r/r_l)$ are regular in $(r/r_l)^{2}$.

The decomposition (\ref{Sbb}) reveals the
inertial-range scaling form of the structure functions.
The leading singular contribution in the limit $r_l\to \infty$, $r_d\to 0$ is given by
the basis operator $\mathcal{F}_\gamma$ with the minimal critical dimension $\Delta_\gamma$
and operators $\mathcal{F}_\alpha$ and $\mathcal{F}_\beta$ with the minimal sum of anomalous dimensions
$\gamma^*_{\alpha}+\gamma^*_{\beta}$. As a result, $S_N$
have singular power-like behavior as $r /r_l \to~0$:
\[
S_{N}(r) \sim r^{N(1-\epsilon/2)} \left({r \over r_d}\right)^{-\gamma^*_{N}}
\left({r \over r_l}\right)^{\Delta_{N}},
\]
with the most negative exponent $\Delta_{N}$ in the basis set generated by the composite
operator $b_r^{N}({\bf x},t)$ and the most negative sum of anomalous dimensions
$\gamma^*_{[M,p]}+\gamma^*_{[K,q]}$ subject to the conditions
$M+K=N$, $p\le M$ and $q\le K$.

In our case, the leading contribution to the sum (\ref{Sbb})
from the OPE (\ref{OPE}) will
be given by the tensor composite operators constructed solely of
the fields ${\bf b}$ without derivatives: $b_{i_1} ... b_{i_p}
(b_i b_i)^l$. It is useful to deal with the scalar operators
obtained by contracting the tensor with the appropriate number of
the anisotropy vectors ${\bf n}$:
\begin{equation}
F[N,p]({\bf x},t) \equiv [{\bf n} \cdot {\bf b}({\bf x},t)]^{p} [b_i({\bf x},t) b_i({\bf x},t)
]^{l}  \label{CO}
\end{equation}
with $N \equiv 2l+p$. Power counting and analysis of graphs
show that composite operators (\ref{CO}) for
given $N$ can be mixed only with each other in
renormalization
\[
F[N,p]=\sum_{p'=0}^N Z_{[N,p][N,p']} F^R[N,p']\,,
\]
therefore, the corresponding  renormalization matrix $Z_{[N,p][N',p']}$
is in fact block-diagonal, i.e., $Z_{[N,p][N',p']}=0$ for
$N'\ne N$.

A detailed account
of practical calculation
of the matrix of the renormalization
constants $Z_{[N,p][N,p']}$
(which may be readily extended
to investigation of all related problems)
has been given
in Ref. \cite{AdAnHnNo00} for the advection of a passive scalar,
therefore we will not describe all details of the
determination of renormalization constants in the present vector model, rather we
will discuss its specific features.

It turned out that not only the $\beta$ functions in
the vector and scalar models coincide, but
the one-loop renormalization matrices as well.
This nontrivial fact stems from
the similarities of the mathematical structure of both models.
In the model of scalar advection
\cite{AdAnHnNo00} the composite operators
$\partial_{i_1}\theta ... \partial_{i_p}\theta (\partial_i\theta
\partial_i \theta)^l$ constructed solely of the scalar gradients
of the scalar admixture $\theta$
are needed for calculation of the asymptotic behavior of the
structure functions, whereas
in our vector case the main contribution is given by
composite operators constructed solely of the fields ${\bf b}$
without derivatives. As direct inspection of the
relevant diagrams shows, the tensor structures arising upon
functional averaging in both cases
are in fact identical, which yields the same renormalization matrix
$Z_{[N,p][N,p']}$ in both models. Thus, it is not necessary to carry out complete
calculations here.

However, in Ref. \cite{AdAnHnNo00} in the expressions for the general
elements of the renormalization matrix of the composite operators there
are misprints (for instance, in the definition of the quantity $Q_1$ in
Eq. (76) of Ref. \cite{AdAnHnNo00} $H_4-H_6$ should be replaced by
$H_3-H_6$), although the numerical investigation of the critical
dimensions is correct. Therefore, we present here the full
formulae, in a slightly different form, however.

The only nonzero elements of the matrix $Z_{[N,p][N,p']}$ are
\begin{align*}
Z_{[N,p][N,p-2]}&=\frac{gC_d}{d^2-1}\frac{1}{4\epsilon}Q_1\,,
&Z_{[N,p][N,p]}&=1+\frac{gC_d}{d^2-1}\frac{1}{4\epsilon}Q_2\,,  \\
Z_{[N,p][N,p+2]}&=\frac{gC_d}{d^2-1}\frac{1}{4\epsilon}Q_3\,,
&Z_{[N,p][N,p+4]}&=\frac{gC_d}{d^2-1}\frac{1}{4\epsilon}Q_4\,,
\end{align*}
with the coefficients $Q_i$ defined as follows
\begin{align*}
Q_1&=  p(p-1)(d+1)( H_0 (1+\alpha_2) +
          (H_4-H_2) (1-2\alpha_1+3 \alpha_2) +
          H_6 (\alpha_1-\alpha_2))\, \\
Q_2&= H_0 ((d-2) (N-p)^2-(1+\alpha_2) (d+1) (p-1) p  \nonumber
      \\
      &\qquad + (N-p) (3+d^2+ 2 d (p-1)+2 p+\alpha_2 (d+1) (1+2 p)))
         \nonumber \\
     &+ H_2 ((5+3 \alpha_2+\alpha_1 (d-2)-d) (N-p)^2 \nonumber \\
     &\qquad +(d+1) (1-\alpha_1+d+\alpha_2 (d+2)) (p-1) p-(N-p)
         (9-2 d+d^2+8 p(d+1) \nonumber \\
      &\qquad - \alpha_1 (3+d^2+2 d (p-1)+2 p)+2 \alpha_2 (4+d+5
      p(d+1)))) \nonumber \\
      &+ H_4 (-(3+6 \alpha_2+\alpha_1 (d-5)) (N-p)^2+(d+1) (
      \alpha_1(d+1)-\alpha_2(2d+1)-d) (p-1)p \nonumber \\
      &\qquad + (N-p) (6 (1+p(d+1))+\alpha_2 (13+d+14 p(d+1))\nonumber \\
      &\qquad- \alpha_1 (9+d^2+8 p+2 d (4p-1))))\nonumber \\
      &+ H_6 (\alpha_2+\alpha_1) (3 (N-p)^2+d (d+1) (p-1) p-6 (N-p)
      (1+p(d+1)))\,,    \\
Q_3&= H_0 (n-p) (-9-d^2+5 N-7 p-d (-2+N+p)+\alpha_2 (d+1) (n-3
      (1+p)))\nonumber \\
      &+ H_2 (N-p) (d^3+d^2 (-2+N+p)+6 (6-3 N+4 p)+d (9-5 N+13
      p)\nonumber \\
        &\qquad +\alpha_2(27-13 N+21 p+d^2 (1+2 p)+d (16-7 N+17
        p))\nonumber \\
        &\qquad +\alpha_1(9+d^2-5 N+7 p+d (-2+N+p)))\nonumber \\
      &+ H_4 (N-p) (-2 (2+d) (6-3 N+(4+d) p) \nonumber \\
       &\qquad+ \alpha_1 (d^3+d^2 (-2+N+p)+6 (6-3 N+4 p)+d (9-5 N+13 p))\nonumber \\
       &\qquad+ \alpha_2(48-24 N+34 p+d^2 (1+4 p)+d (25-12 N+26 p)))
       \nonumber \\
       &+ H_6 2(N-p) (\alpha_2-\alpha_1) (2+d) (6-3 N+(4+d) p)\,,
       \\
Q_4&= H_0 (3+\alpha_2(d+1)) (2-N+p) (N-p)\nonumber \\
       &+ H_2 (3 \alpha_1-6 (2+d)-\alpha_2 (9+7 d+d^2)) (2-N+p)
       (N-p)\nonumber \\
       &+ H_4 (2+d) (-6 \alpha_1+(1+2\alpha_2) (4+d)) (2-N+p)
       (N-p)\nonumber \\
       &+ H_6 (\alpha_1-\alpha_2) (8+6 d+d^2) (2-N+p) (N-p) \,,
\end{align*}
where $H_i$ are the functions
\begin{align*}
H_0&= {_2F_1}(1,1/2,d/2,-\chi)\,, \\
H_2&=   {_2F_1}(1,3/2,d/2+1,-\chi)/d\,, \\
H_4&= {_2F_1}(1,5/2,d/2+2,-\chi) \frac{3}{d(d+2)}\,, \\
H_6&=  {_2F_1}(1,7/2,d/2+3,-\chi) \frac{15}{d(d+2)(d+4)}\,.
\end{align*}
Here, $_2F_1$ is the Gauss hypergeometric function.
The anomalous dimensions $\gamma_{[N,p][N,p']}$ are
\begin{align}
\gamma_{[N,p][N,p-2]}&=-\frac{gC_d}{4(d^2-1)}Q_1\,,
&\gamma_{[N,p][N,p]}&=-\frac{gC_d}{4(d^2-1)}Q_2\,, \nonumber \\
\gamma_{[N,p][N,p+2]}&=-\frac{gC_d}{4(d^2-1)}Q_3\,,
&\gamma_{[N,p][N,p+4]}&=-\frac{gC_d}{4(d^2-1)}Q_4\,, \label{andim}
\end{align}
and the matrix of critical dimensions (\ref{deltaf}) is
\begin{equation}
\Delta_{[N,p][N,p']}=-N\left(1-{\epsilon\over 2}\right)\delta_{pp'} +
\gamma^*_{[N,p][N,p']}\,,\label{deltaff}
\end{equation}
where the asterisk stands for the value at the fixed point
(\ref{fixg}), (\ref{fixchi}). This represents the critical dimensions
of the composite operators (\ref{CO}) at the first order in
$\epsilon$. It should to be stressed that in contrast to the value of
the fixed point (\ref{fixg}), (\ref{fixchi}), which has no higher
order corrections, the expressions for anomalous dimensions
(\ref{andim}) have nonvanishing corrections of order $g^2$ and higher.

The critical dimensions are given by the eigenvalues of the matrix
(\ref{deltaff}). As was already discussed in
Ref. \cite{AdAnHnNo00} in the limiting isotropic case
($\alpha_{1}=\alpha_{2}=0$) this matrix becomes triangular, i.e., the
eigenvalues are simply equal to the diagonal elements $\Delta[N,p]
\equiv \Delta_{[N,p][N,p]}$.

Since our result for the anomalous dimensions is the same as in
Ref. \cite{AdAnHnNo00} for the admixture of a passive scalar, all
conclusions about the hierarchical behavior of the critical
dimensions of the composite operators are also valid in the
analysis of the present model. Nevertheless, the inertial-range asymptotic behavior
of the structure functions
in these two problems is completely different, because, first, in the scalar problem
single-point products of the scalar are not renormalized, while in the vector problem
they are, and, second, the leading contribution to the OPE is given by the products of derivatives
of the scalar, whereas in the vector problem products of the field components themselves
yield the leading contribution.

In Ref. \cite{AdAnHnNo00}
the behavior of the critical dimensions
$\Delta[N,p]$ for $N=2,3,4,5,$ and $6$ was
numerically studied.  The main conclusion is
that the dimensions $\Delta_N$ remain negative in anisotropic case
and decrease monotonically as $N$ increases for odd and even
values of $N$ separately.

In present paper we concentrate our attention on the investigation
of the composite operators (\ref{CO}) for relatively large values
of $N$, namely we will analyze cases with
$N=10,11,20,21,30,31,40,41,50$, and $51$. Our aim has been to
find out whether hierarchies which hold for small values of $N$
remain valid for significantly larger values of $N$, and the answer turned out to be
in the affirmative.

\begin{figure}
       \includegraphics[width=5.5cm]{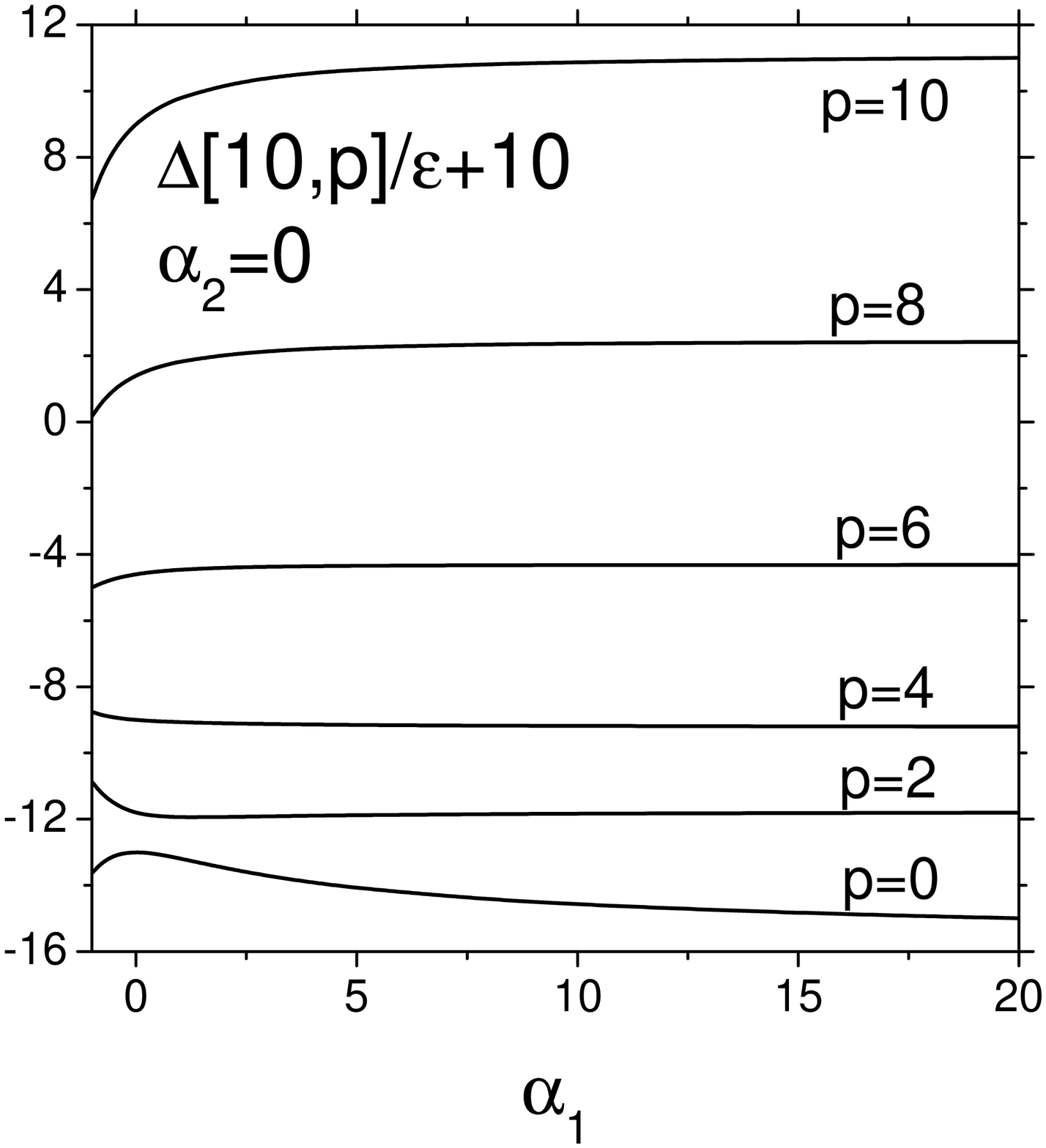}
       \includegraphics[width=5.5cm]{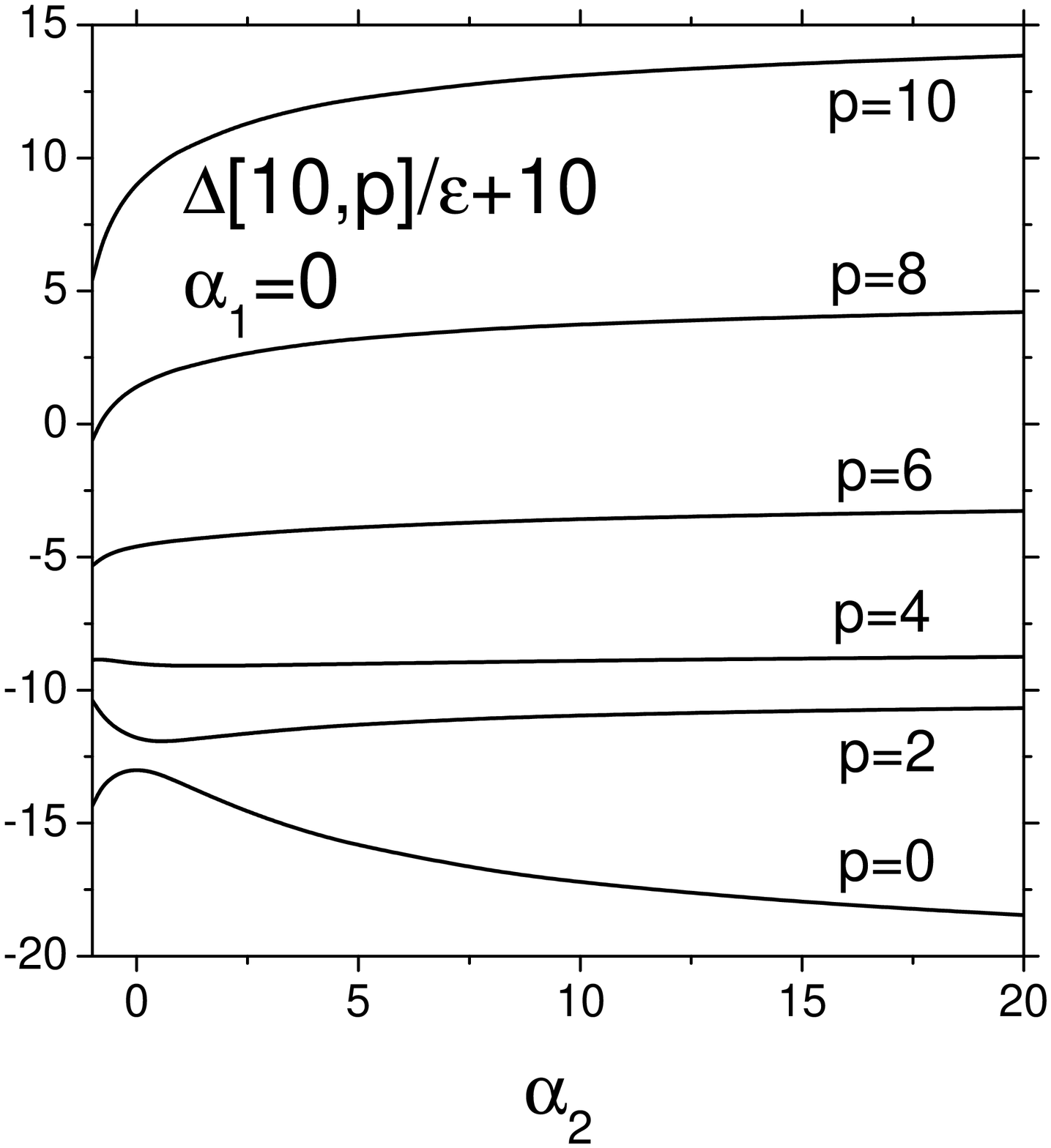}
       \includegraphics[width=5.5cm]{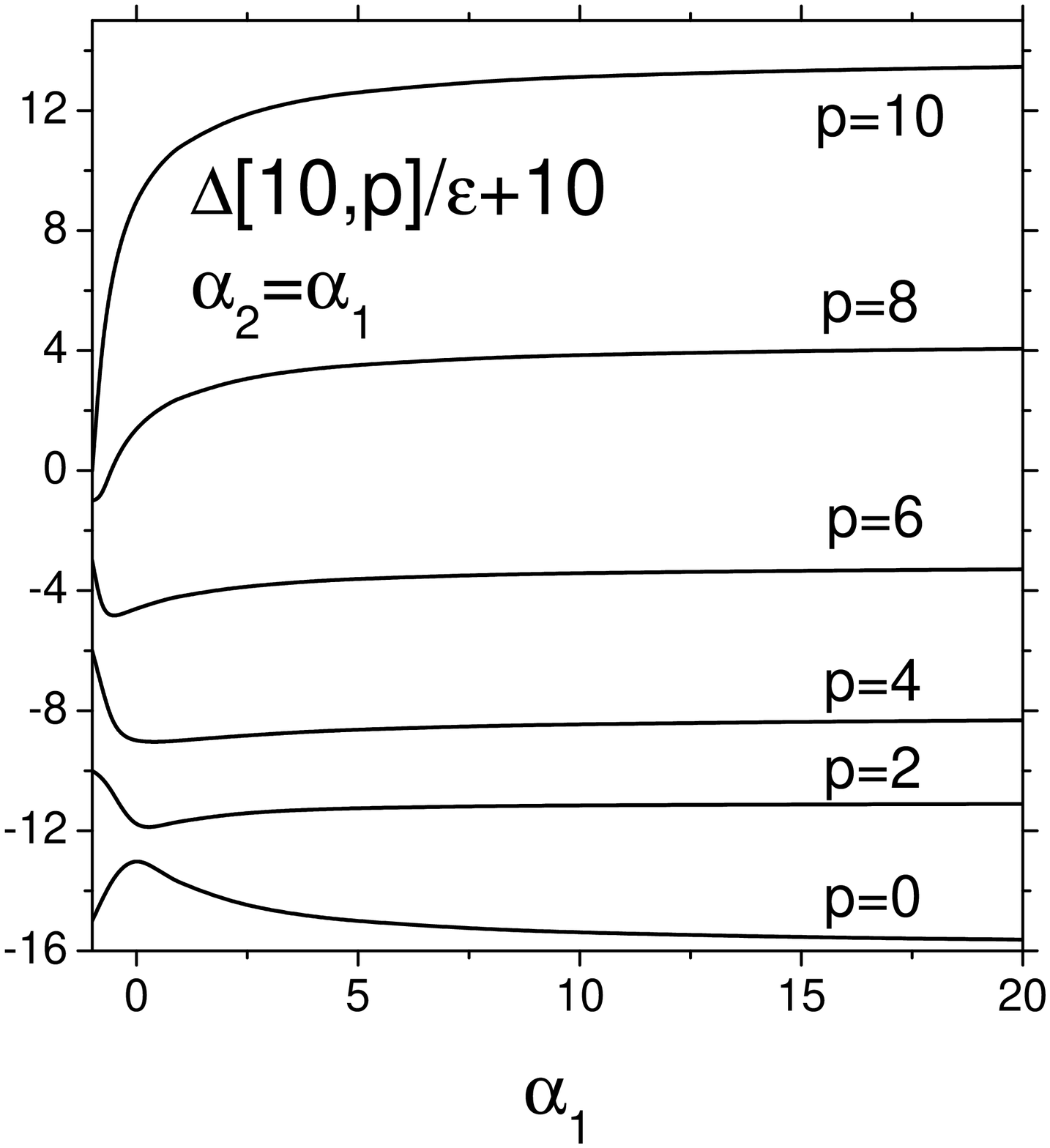}
\caption{Behavior of the critical dimension
$\Delta[10,p]/\epsilon$ for space dimension $d=3$ and for
representative values of $p$ as functions of anisotropy parameters
$\alpha_1$ and $\alpha_2$.  \label{fig1}}
\end{figure}

\begin{figure}
       \includegraphics[width=5.5cm]{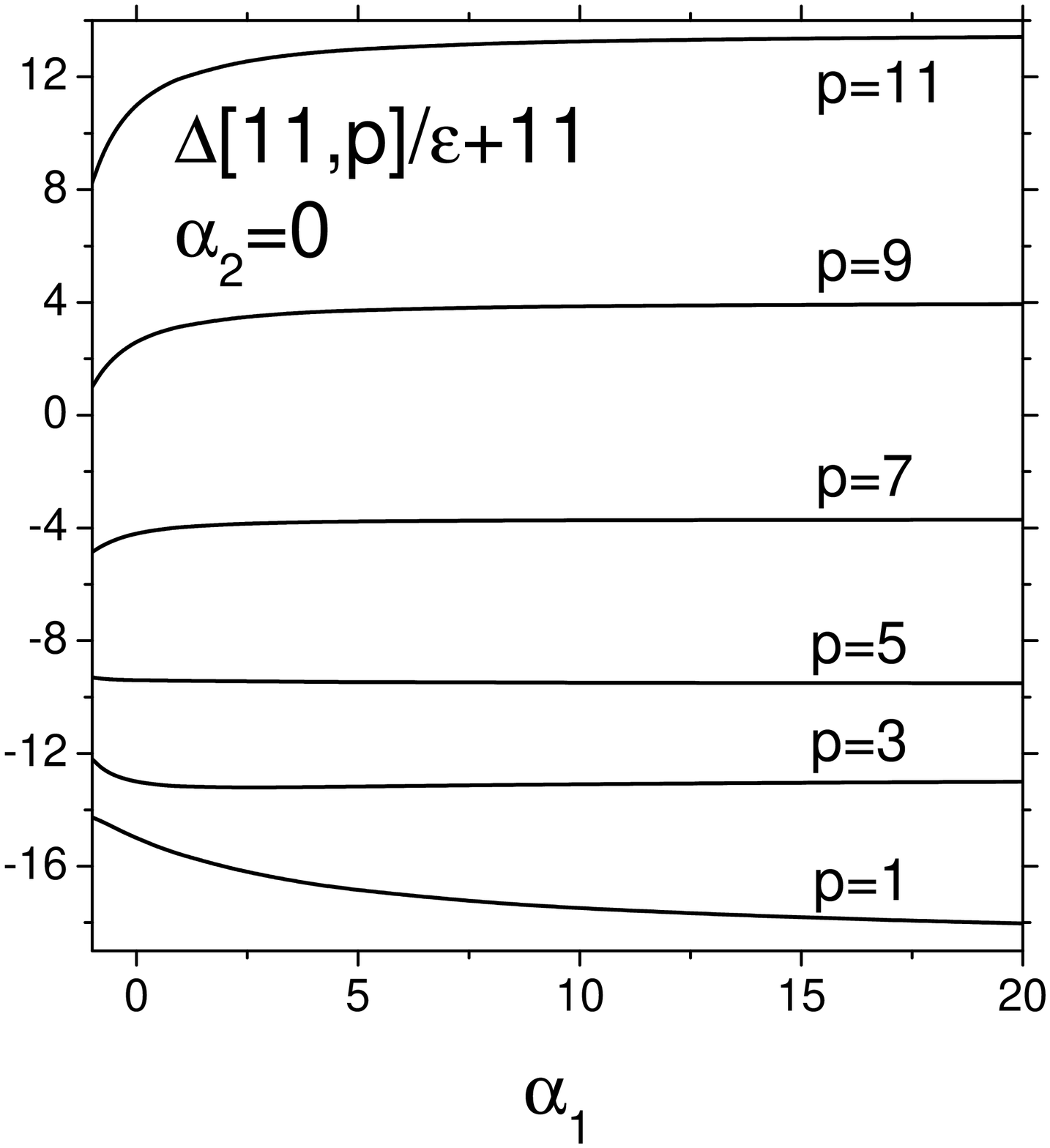}
       \includegraphics[width=5.5cm]{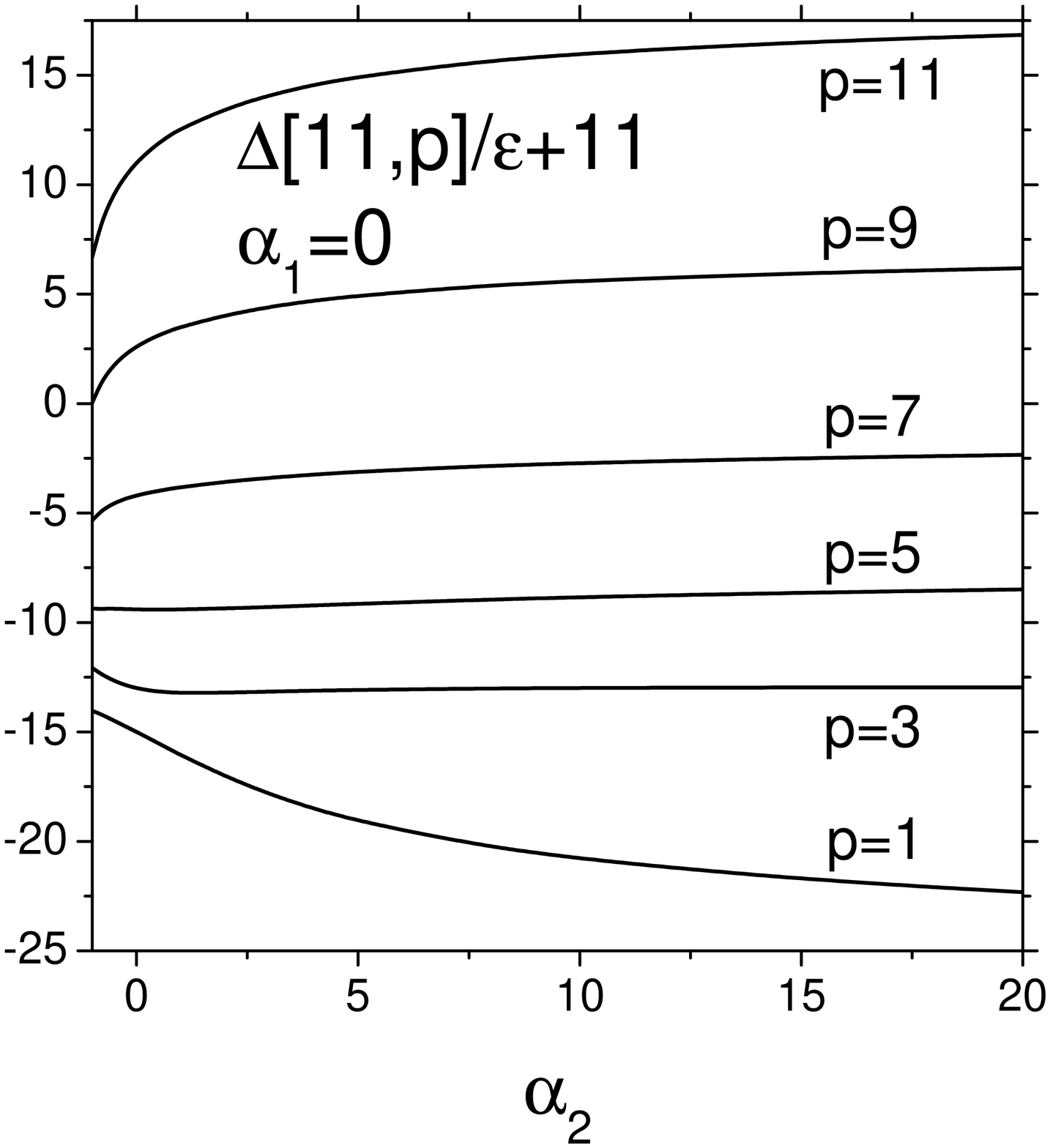}
       \includegraphics[width=5.5cm]{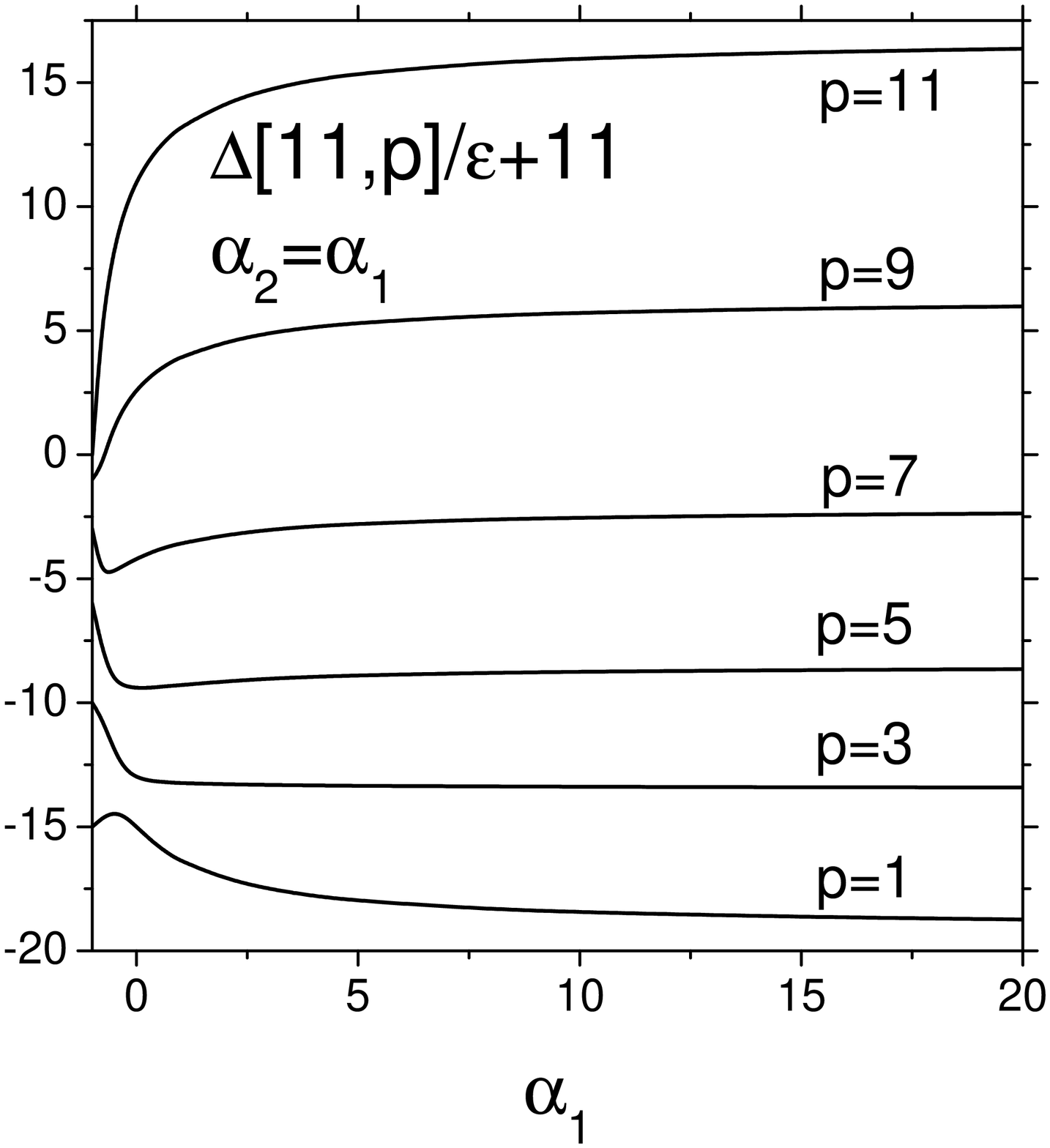}
\caption{Behavior of the critical dimension
$\Delta[11,p]/\epsilon$ for space dimension $d=3$ and for
representative values of $p$ as functions of anisotropy parameters
$\alpha_1$ and $\alpha_2$.  \label{fig2}}
\end{figure}

\begin{figure}
       \includegraphics[width=5.5cm]{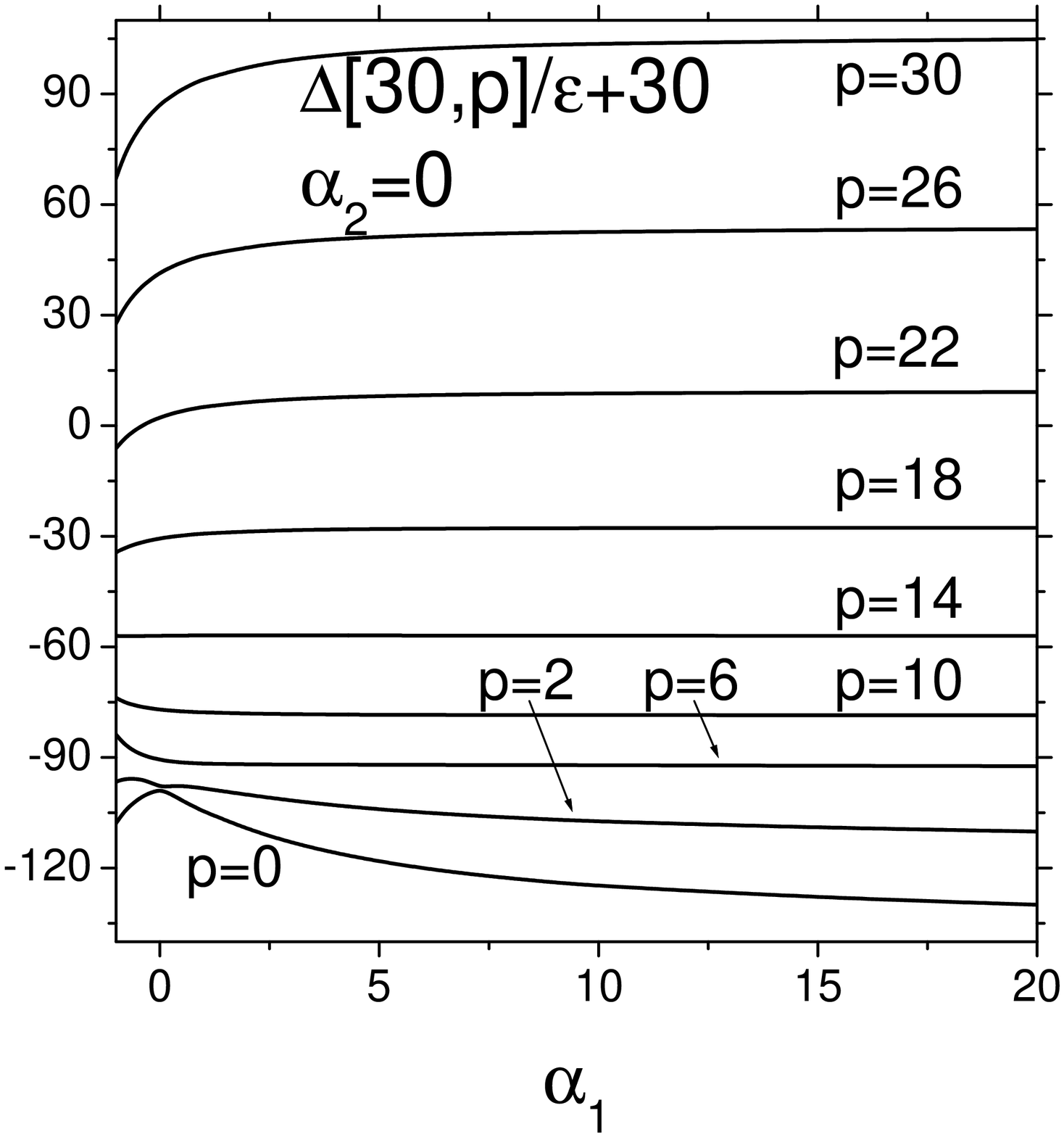}
       \includegraphics[width=5.5cm]{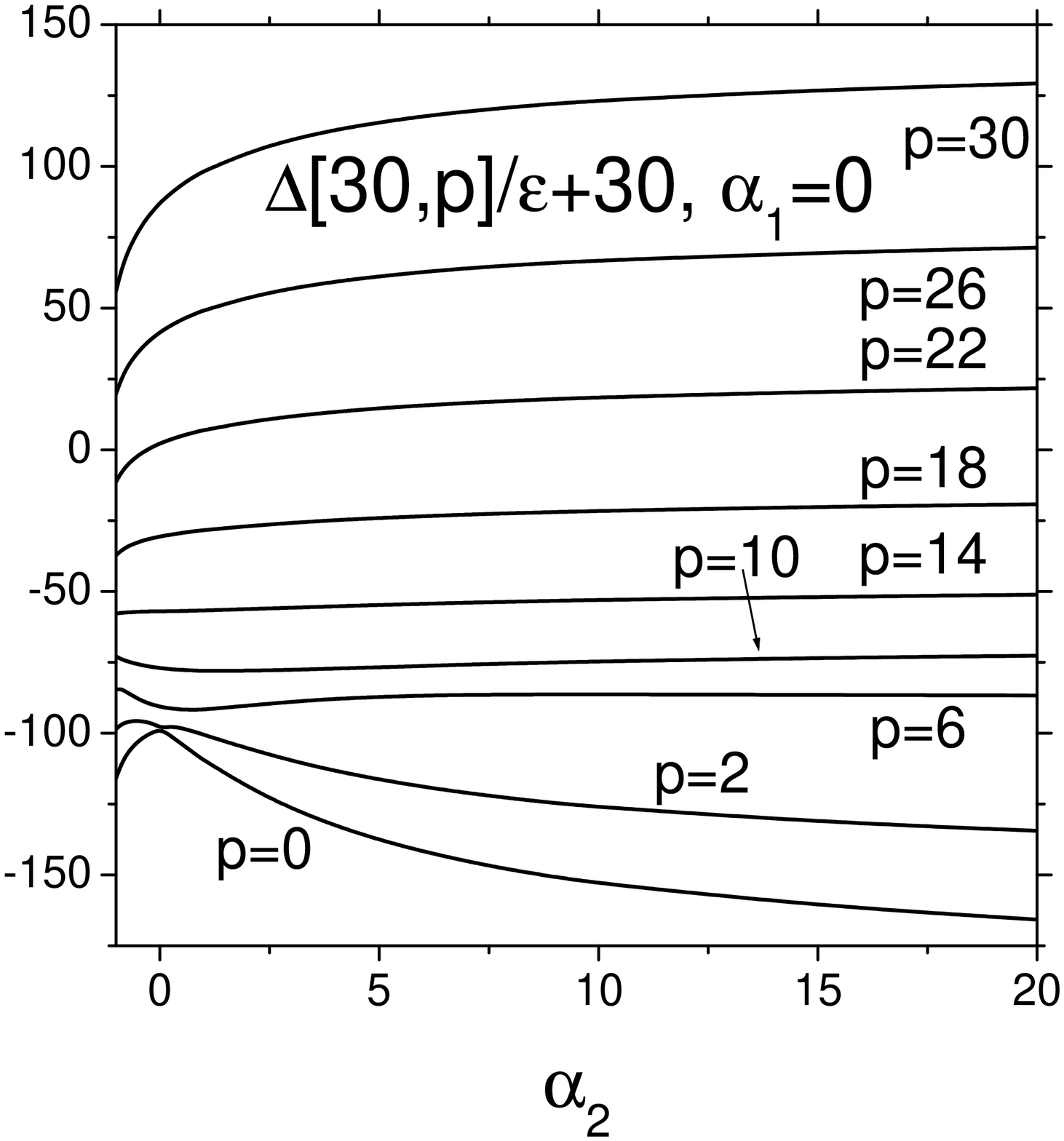}
       \includegraphics[width=5.5cm]{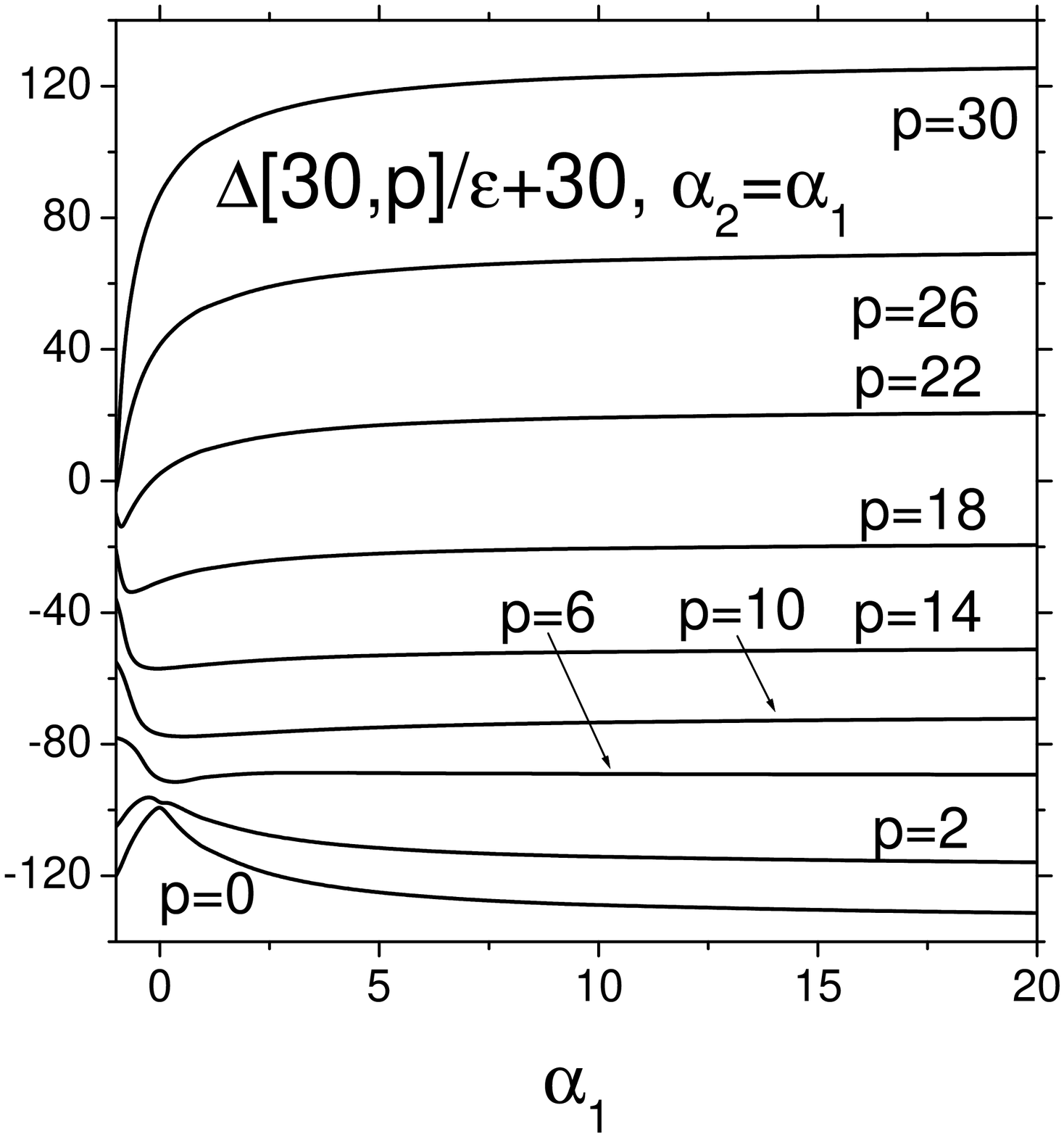}
\caption{Behavior of the critical dimension
$\Delta[30,p]/\epsilon$ for space dimension $d=3$ and for
representative values of $p$ as functions of anisotropy parameters
$\alpha_1$ and $\alpha_2$.  \label{fig3}}
\end{figure}

\begin{figure}
       \includegraphics[width=5.5cm]{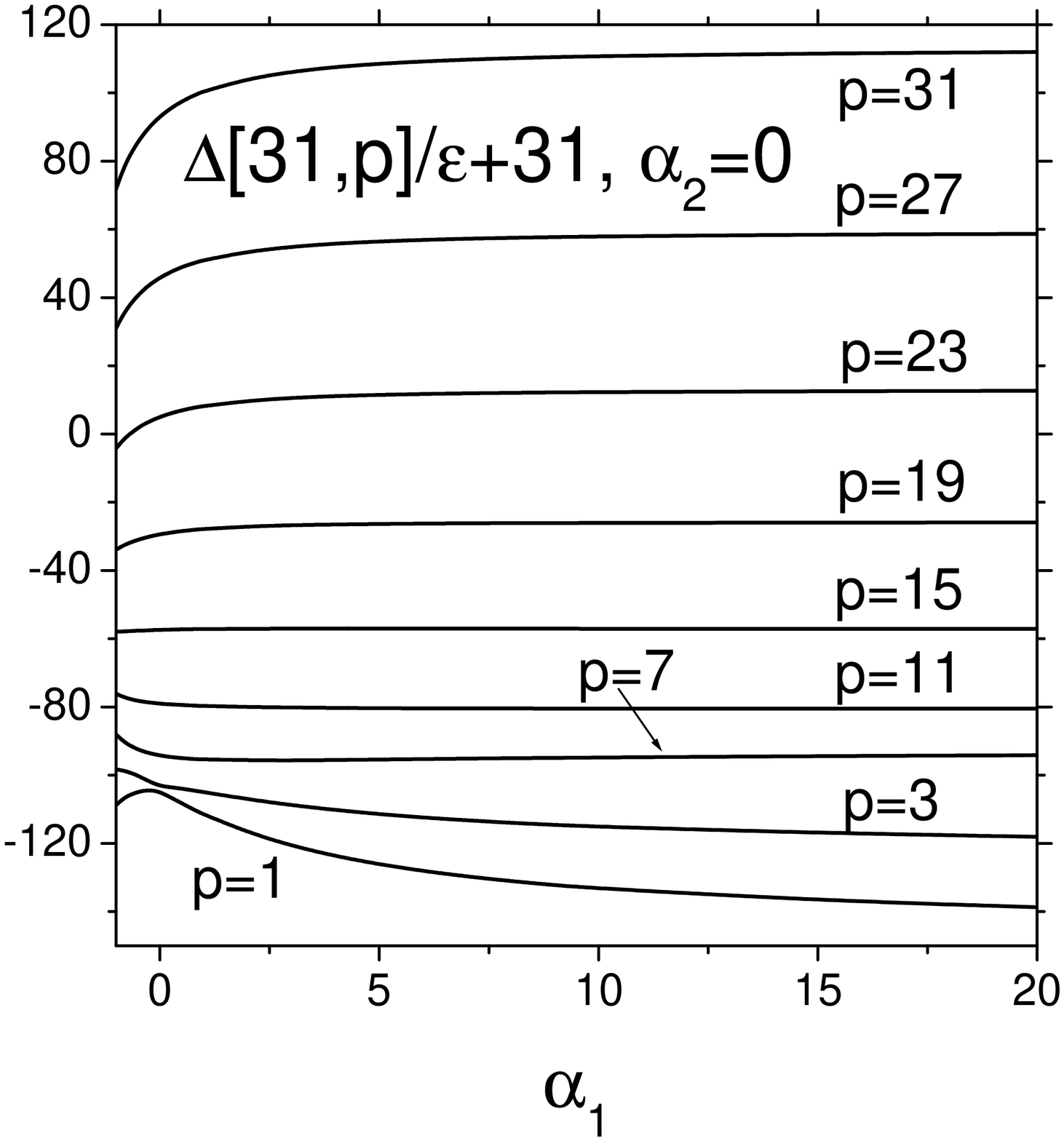}
       \includegraphics[width=5.5cm]{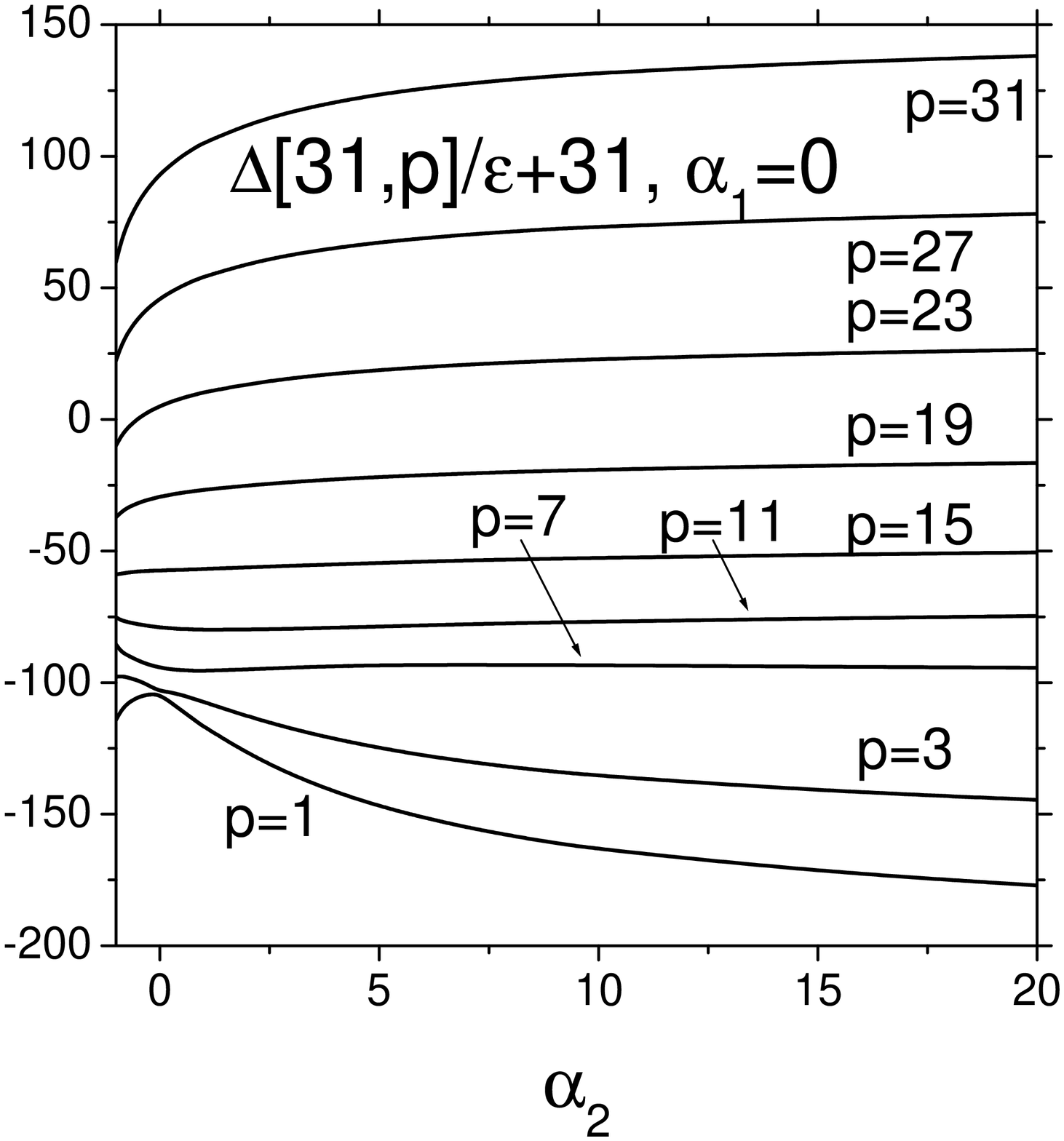}
       \includegraphics[width=5.5cm]{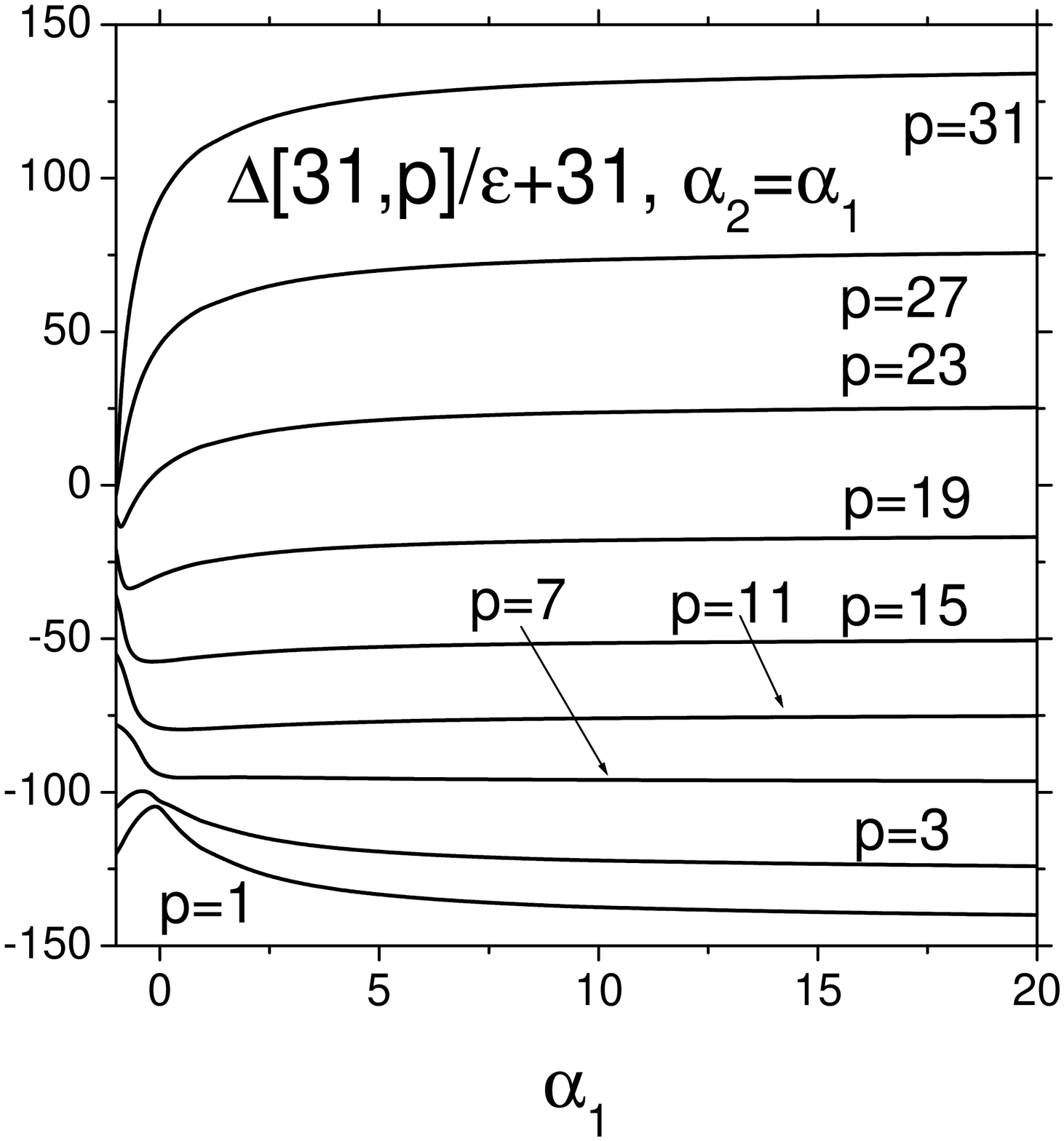}
\caption{Behavior of the critical dimension
$\Delta[31,p]/\epsilon$ for space dimension $d=3$ and for
representative values of $p$ as functions of anisotropy parameters
$\alpha_1$ and $\alpha_2$.  \label{fig4}}
\end{figure}

\begin{figure}
       \includegraphics[width=5.5cm]{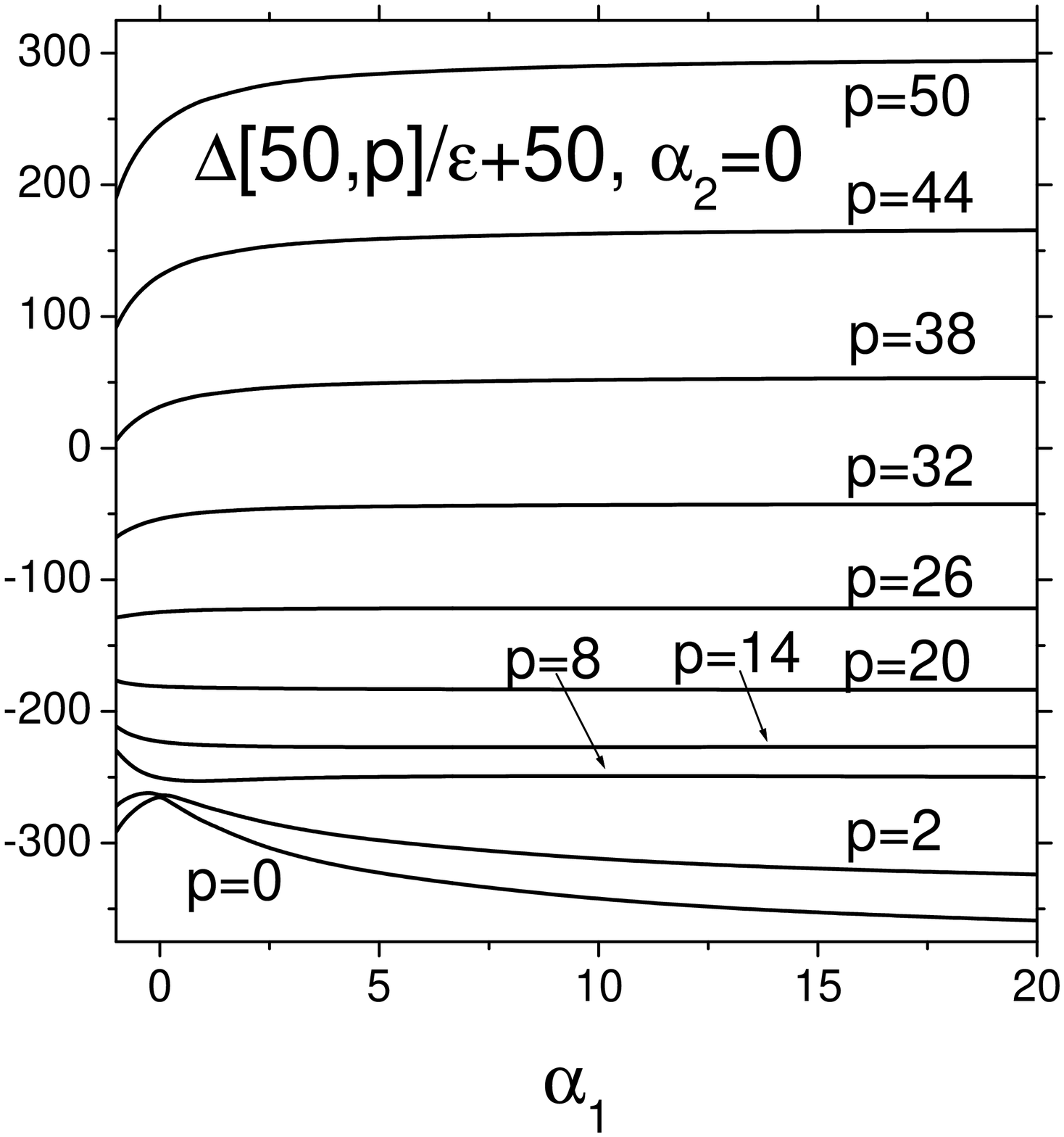}
       \includegraphics[width=5.5cm]{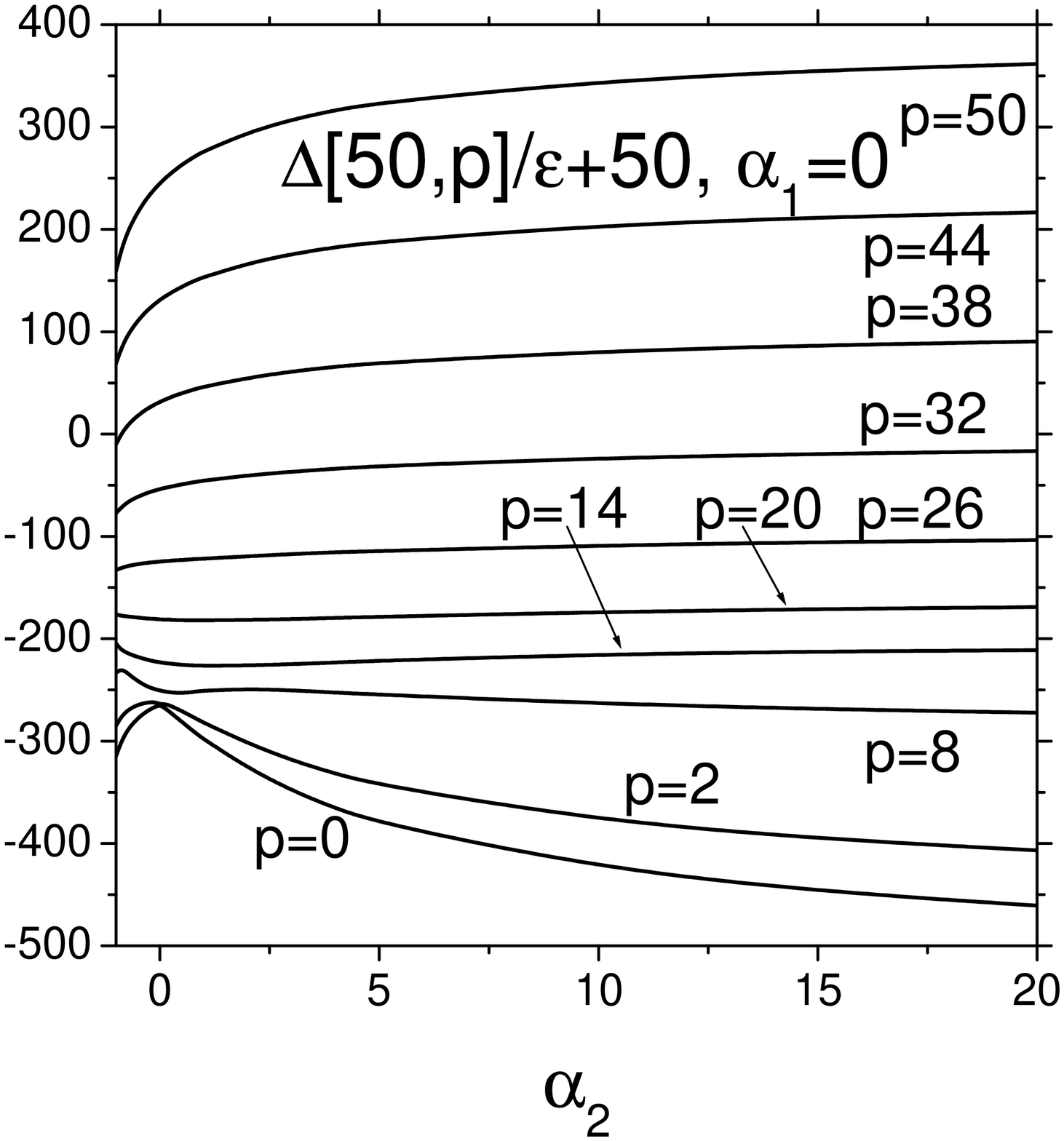}
       \includegraphics[width=5.5cm]{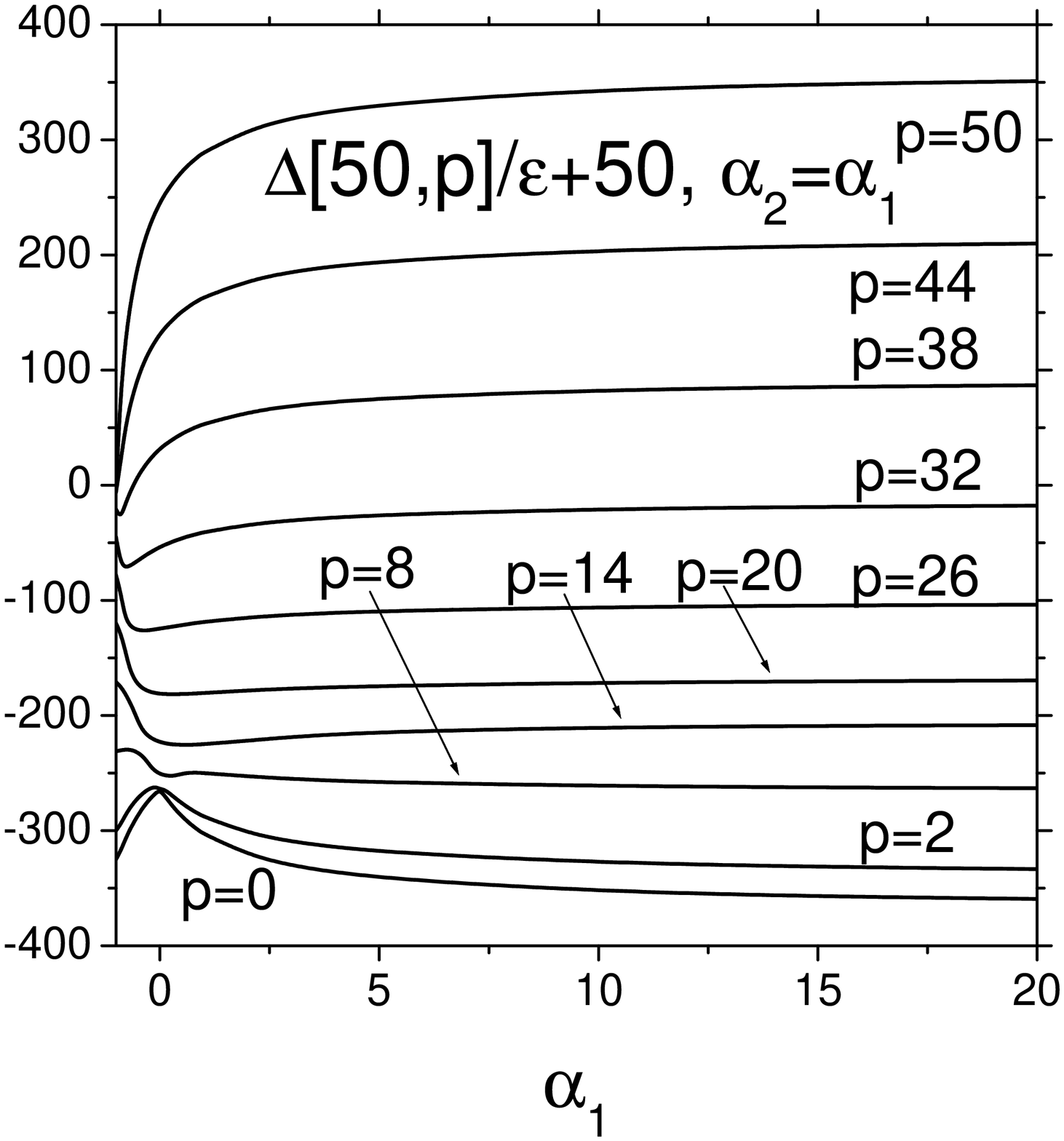}
\caption{Behavior of the critical dimension
$\Delta[50,p]/\epsilon$ for space dimension $d=3$ and for
representative values of $p$ as functions of anisotropy parameters
$\alpha_1$ and $\alpha_2$.  \label{fig5}}
\end{figure}

\begin{figure}
       \includegraphics[width=5.5cm]{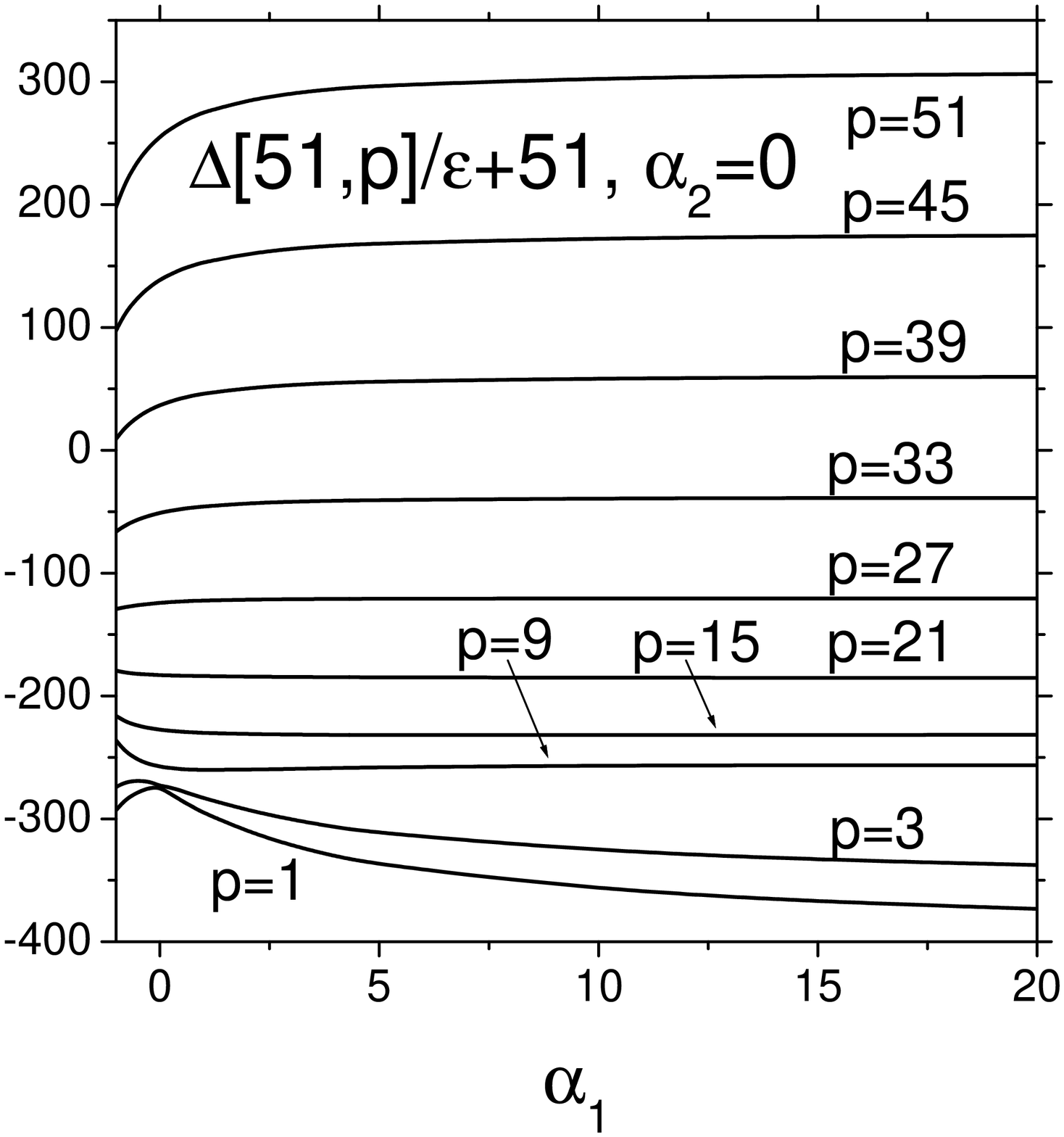}
       \includegraphics[width=5.5cm]{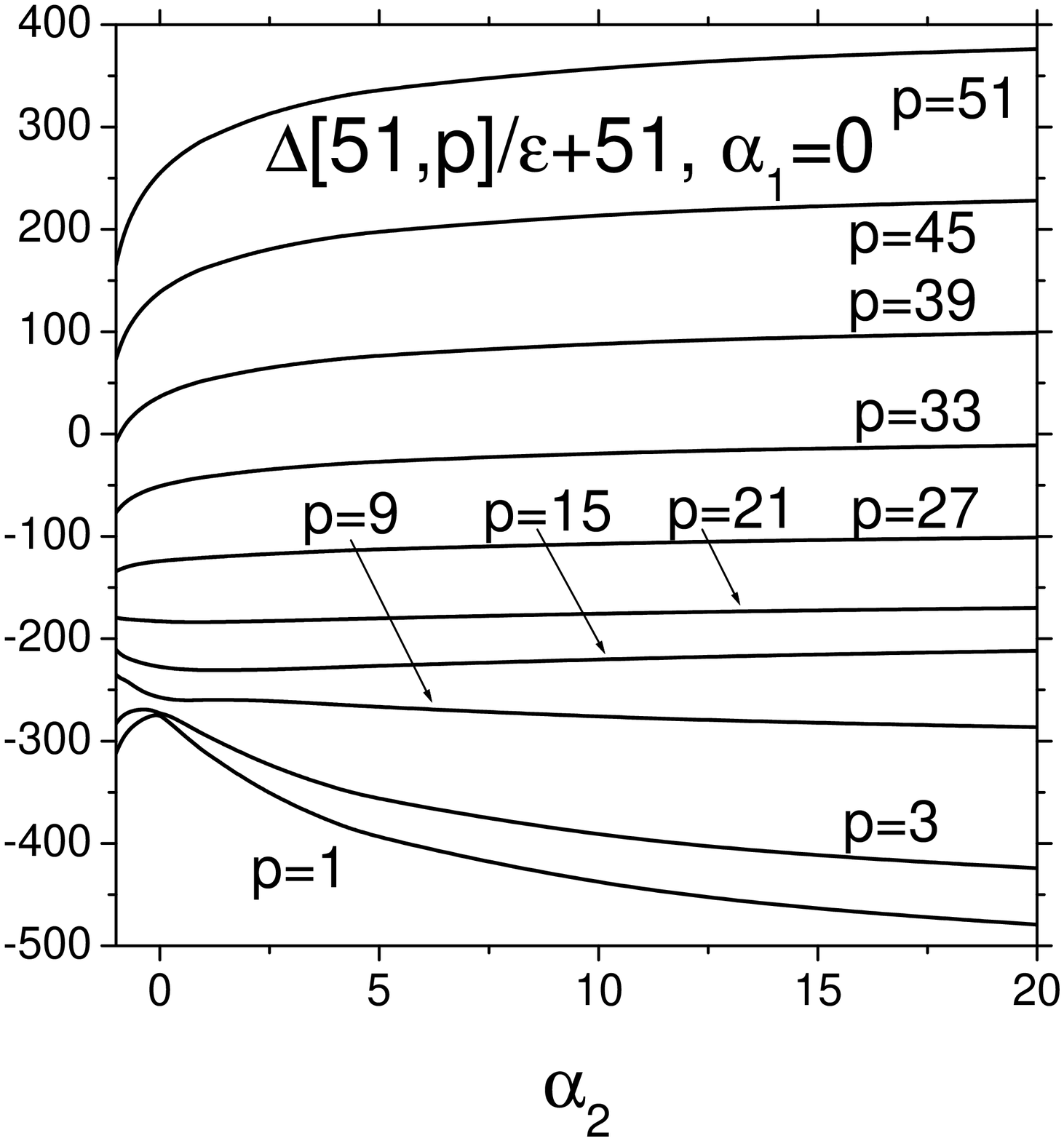}
       \includegraphics[width=5.5cm]{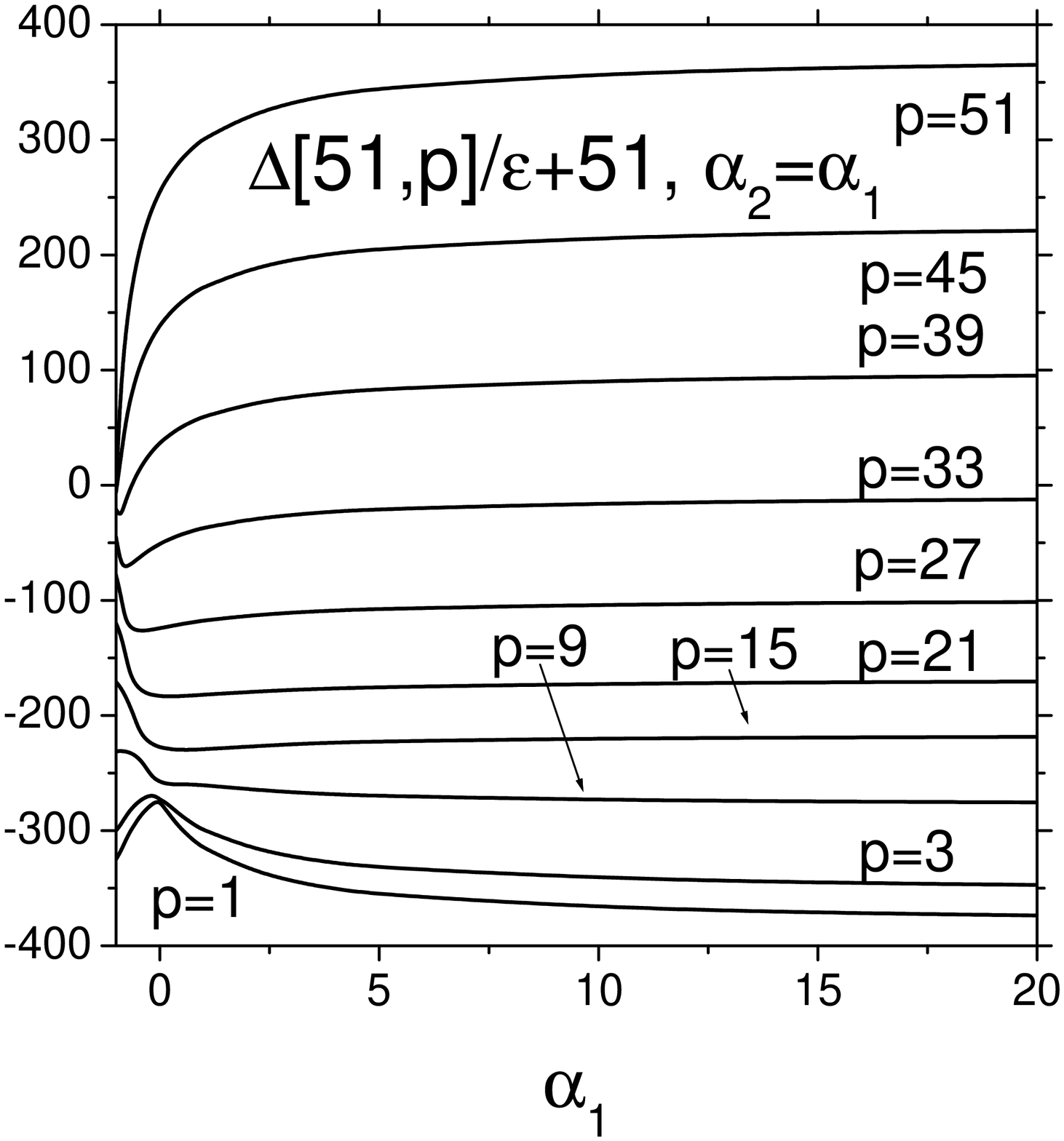}
\caption{Behavior of the critical dimension
$\Delta[51,p]/\epsilon$ for space dimension $d=3$ and for
representative values of $p$ as functions of anisotropy parameters
$\alpha_1$ and $\alpha_2$.  \label{fig6}}
\end{figure}

In Ref. \cite{AdAnHnNo00} several hypothetically
possible structures of the matrix of critical dimensions
(\ref{deltaff}) were discussed.
In particular,  the possibility that the matrix
(\ref{deltaff}) for some $\alpha_{1}$ and $\alpha_{2}$ would have a pair of complex
conjugate eigenvalues $\Delta = {\rm Re}\,\Delta \pm i\, {\rm Im}\, \Delta $
cannot be excluded {\em a priori}. In
this case, the small-scale behavior of the scaling functions would
have oscillating terms of the form
\[
\left({r\over r_l}\right)^{{\rm Re}\, \Delta}
\left\{C_1 \cos\left[({\rm Im}\, \Delta) r/r_l\right]+C_2
\sin\left[({\rm Im}\,\Delta)r/r_l\right]\right\}\,,
\]
with some real constants $C_{1}$, $C_{2}$.

Another, in general, conceivable
structure of the matrix (\ref{deltaff}) is related to the situation
when it cannot be diagonalized but only reduced to the
Jordan normal form. In this case, the corresponding contribution to the
scaling function would involve a logarithmic correction to the
power-like behavior, viz.
\[
\left({r\over r_l}\right)^{\Delta} [C_1 \ln(r/r_l)+C_2]\,,
\]
where $\Delta$ is the eigenvalue related to the Jordan cell.

In Figs. \ref{fig1}-\ref{fig6}  behavior of the
eigenvalues of the matrix of critical dimensions $\Delta[N,p]$ for
relatively large values of the $N$ are shown. It can be seen that only real
eigenvalues exist in all cases, and also their hierarchical
behavior discussed in Ref. \cite{AdAnHnNo00} is conserved. At
first sight the curves for $p=0$ and $p=2$ in the even case and
the curves for $p=1$ and $p=3$ in the odd case in
Figs. \ref{fig3}-\ref{fig6} appear to be crossing at the point $\alpha_1 =
\alpha_2=0$ but in fact the curves are only visually
running very near together at that point which is a mathematical
consequence of the formulas for critical dimensions in the
infinitesimal limit $\alpha_1 \rightarrow 0$ and $\alpha_2
\rightarrow 0$.

%\begin{figure}
%       \includegraphics[width=5.5cm]{fig8a.eps}\hspace{3cm}
%       \includegraphics[width=5.5cm]{fig8b.eps}
%\caption{Behavior of the critical dimensions
%$\Delta[30,0]/\epsilon$ and $\Delta[31,1]/\epsilon$ for space
%dimension $d=3$ as functions of anisotropy parameters $\alpha_1$
%and $\alpha_2$.  \label{fig7}}
%\end{figure}

%\begin{figure}
%       \includegraphics[width=5.5cm]{fig9a.eps}\hspace{3cm}
%       \includegraphics[width=5.5cm]{fig9b.eps}
%\caption{Behavior of the critical dimensions
%$\Delta[50,0]/\epsilon$ and $\Delta[51,1]/\epsilon$ for space
%dimension $d=3$ as functions of anisotropy parameters $\alpha_1$
%and $\alpha_2$.  \label{fig8}}
%\end{figure}

%In Figs. \ref{fig7}, \ref{fig8} the eigenvalues are
%presented as the functions of two variables $\alpha_1$ and
%$\alpha_2$ for $N=30,31,50,51$ for the first the most singular modes
%$p=0$ for even $N$ and $p=1$ for odd ones. It can clearly be seen
%that as values of the parameters $\alpha_{1}$ and $\alpha_{2}$
%increase the critical dimensions become more negative approaching
%some saturated values, therefore the anisotropy amplifies the
%anomalous scaling.

\section{Conclusions}

In this paper we have analyzed asymptotic behavior of the structure
functions $S_N$ of passively advected vector field with small-scale
anisotropy. To this end field-theoretic renormalization group
and the operator-product expansion have been used in a minimal-subtraction
scheme of analytic renormalization.

It is shown that the leading-order asymptotic behavior of the structure functions
is determined by the isotropic sector of the velocity field.
At the leading order in the inertial interval all the structure functions
are  flat, i.e., independent of the separation distance, with powerlike
corrections (with real positive exponents) effected by
the small-scale anisotropy.
We have calculated numerically the anomalous
correction exponents up to order $N=51$ to explore possible oscillatory
modulation
or logarithmic corrections to the leading powerlike asymptotics, but have
found no sign of this kind of behavior: all calculated corrections have
had purely powerlike behavior.
Our results show that the exponents of the powerlike corrections tend
to grow with increasing relative impact of the anisotropy.

From the renormalization-group point of view the present model of
passively advected vector field in the presence of
strong anisotropy has turned out to be very similar to
that of passively advected scalar field \cite{AdAnHnNo00}. In
particular, the $\beta$ functions and the one-loop
contributions to renormalization matrices of relevant
composite operators are the same. Since
in the published analysis of the scalar problem \cite{AdAnHnNo00} there were some misprints,
we have also presented corrected complete results of the calculation of the
renormalization matrices.

However, physically the two models differ significantly: instead of the
anomalous powerlike growth of the structure functions of the scalar problem
in the inertial range, in the present
vector case the leading asymptotic term turned out to be constant
with powerlike corrections due to the small-scale anisotropy.
Thus, for the passively advected
vector field a much stronger departure from the Kolmogorov-like
scaling than in the case of passively advected scalar is predicted.

\acknowledgements

M. H. gratefully acknowledges the hospitality of the N. N. Bogoliubov
Laboratory of Theoretical Physics at JINR, Dubna, the University of Genova, Italy, and the Department of
Physical Sciences
of the University of Helsinki, Finland. J. H. thanks the Institute for Experimental
Physics of the Slovak Academy of Sciences in Ko\v{s}ice for
hospitality.

This work was supported in part by the Slovak Academy of Sciences (VEGA Grant No. 3211),
by the Academy of Finland (Grant No. 203121), and by the common grant CNR (Italy) and SAS (Slovakia).

\end{document}